\documentclass[10pt,journal,compsoc]{IEEEtran}

\ifCLASSOPTIONcompsoc
  \usepackage{cite}
\else
  \usepackage{cite}
\fi
\ifCLASSINFOpdf
\else
\fi
\usepackage{CJK}
\usepackage{epstopdf}
\usepackage{subfigure}
\usepackage{color}
\usepackage{ulem}
\usepackage{CJKulem}
\usepackage{caption}
\usepackage{graphicx, subfig}
\usepackage{amsmath,bm}
\usepackage{amsmath,amssymb}

\usepackage{booktabs}
\usepackage{float}
\usepackage{multirow}
\usepackage{pifont}
\usepackage{bbding}
\usepackage{makecell}
\usepackage{ulem}
\usepackage{url}
\usepackage{bm}

\usepackage{draftwatermark, everypage}
\SetWatermarkText{Draft}
\SetWatermarkLightness{0.85} 
\SetWatermarkScale{1.3}

\usepackage[dvipsnames, table]{xcolor}
\definecolor{c1}{HTML}{6e9a44}
\definecolor{c2}{HTML}{67a030}
\definecolor{c3}{HTML}{4e7851}

\usepackage{stfloats}
\usepackage{enumitem}
\usepackage{colortbl}
\definecolor{midorange}{RGB}{251,235,211}
\definecolor{lightorange}{RGB}{253,243,231}

\usepackage{stackengine}

\usepackage{url}

\begin{document}
%
\title{GotFlow3D: Recurrent Graph Optimal Transport for Learning 3D Flow Motion in Particle Tracking}
%
%
%
%

\author{Jiaming~Liang,
	Chao~Xu
	and~Shengze~Cai
	\IEEEcompsocitemizethanks{\IEEEcompsocthanksitem J. Liang and C. Xu are affiliated with the College of Control Science \& Engineering, Zhejiang University (ZJU), Hangzhou, 310027 China, the State Key Laboratory of Industrial Control Technology, the National Engineering Center for Industrial Automation, the Institute of Cyber-Systems \& Control at ZJU, and ZJU Huzhou Institute.\protect\\
 S. Cai is with the Institute of Cyber-Systems \& Control, College of Control Science \& Engineering, Zhejiang University, Hangzhou, 310027 China. \protect\\
Corresponding authors: shengze\_cai@zju.edu.edu, cxu@zju.edu.cn
}
}

%
%

\markboth{Journal of \LaTeX\ Class Files}
{Shell \MakeLowercase{\textit{et al.}}: Bare Advanced Demo of IEEEtran.cls for IEEE Computer Society Journals}
%



\IEEEtitleabstractindextext{%
\begin{abstract}
Flow visualization technologies such as particle tracking velocimetry (PTV) are broadly used in understanding the all-pervasiveness three-dimensional (3D) turbulent flow from nature and industrial processes.
Despite the advances in 3D acquisition techniques, the developed motion estimation algorithms in particle tracking remain great challenges of large particle displacements, dense particle distributions and high computational cost.
By introducing a novel deep neural network based on recurrent Graph Optimal Transport, called GotFlow3D, we present an end-to-end solution to learn the 3D fluid flow motion from double-frame particle sets. 
The proposed network constructs two graphs in the geometric and feature space and further enriches the original particle representations with the fused intrinsic and extrinsic features learnt from a graph neural network.
The extracted deep features are subsequently utilized to make optimal transport plans indicating the correspondences of particle pairs, which are then iteratively and adaptively retrieved to guide the recurrent flow learning.
Experimental evaluations, including assessments on numerical experiments 
and validations on real-world experiments, demonstrate that the proposed GotFlow3D achieves state-of-the-art performance against both recently-developed scene flow learners and particle tracking algorithms, with impressive accuracy, robustness and generalization ability, which can provide deeper insight into the complex dynamics of broad physical  and biological systems. 
\end{abstract}

\begin{IEEEkeywords}
Graph Neural Networks, Recurrent Neural Networks, Optimal Transport, Fluid Motion Estimation, Point Clouds.
\end{IEEEkeywords}}

\maketitle

\IEEEdisplaynontitleabstractindextext

%
\IEEEpeerreviewmaketitle

\ifCLASSOPTIONcompsoc
\IEEEraisesectionheading{\section{Introduction}\label{sec:introduction}}
\else
\section{Introduction}
\label{sec:introduction}
\fi

\IEEEPARstart{V}{isualization} and measurement of the turbulent flow can reveal dynamic behaviors of great complexity in fluid mechanics, biological locomotion and soft materials, which has long been recognized by Leonardo da Vinci and carried out to track fine particles in turbulent flows of the aortic heart valve in the 1400s \cite{kemp2019leonardo}.
Among the diverse flow visualization technologies, particle tracking velocimetry (PTV) is commonly used for quantitative and non-intrusive measurement of global velocity field, that can provide a deeper perception into the complex flow phenomena \cite{dabiri2020particle}.
In particular, tracking the suspended particles embedded within a fluid medium (e.g., liquid, gas or other material with continuous deformation) in PTV has been playing a significant role in elucidating natural phenomena formation processes \cite{kopitca2021programmable, ferdowsi2017river}, hydrodynamics of biological locomotion \cite{hu2003hydrodynamics}, tissue-scale or intracellular flow in organisms \cite{he2014apical, mestre2020cerebrospinal, zhang2022cerebral, guo2014probing}, biological collective phenomena or transport processes \cite{peng2021imaging, schuerle2019synthetic}, dynamic mechanisms of liquids \cite{punzmann2014generation} as well as atomic rearrangement motion in solids \cite{huang2013imaging}.

The PTV algorithms  rely on the detection and tracking of individual particles between consecutive particle images, resulting in Lagrangian velocity vectors that can considerably represent the local motions of the fluid flow. In general, the particles are detected and localized with sub-pixel spatial coordinates in different ways depending on various experimental setups (e.g., defocusing PTV \cite{pereira2006two} or traction force microscopy  \cite{leggett2020mechanophenotyping}).
Compared to other developed flow analysis techniques, e.g., particle image velocimetry (PIV) \cite{raffel2018particle}, PTV shows the great superiority for providing flow fields with substantially increased spatial resolution as well as improved  accuracy by avoiding bias errors due to the spatial averaging in PIV~\cite{dabiri2020particle}.
Moreover, due to the individual particle tracking manner, PTV demonstrates better performance in measuring the flows with strong velocity gradients and inhomogeneity of particle spatial distribution due to Saffman effect \cite{raffel2018particle}, especially in a three-dimensional (3D) spatial domain. 

Despite the aforementioned advantages, the ever-growing PTV demands in scientific researches and engineering applications pose great challenges to classical particle tracking algorithms, which need to be further generalized to tricky flow scenarios involving  diverse motion patterns, large dynamic velocity ranges,  dense particle distribution, etc. 
In addition, the classical approaches commonly involve complicated iterative algorithms with adjustable parameters or introduce additional temporal smoothness from multiple frames \cite{cierpka2013higher, dabiri2020particle}, which are generally time-consuming in real experiments. 
To overcome these limitations, we incorporate deep neural networks developed for 3D point clouds to solve the fundamental problem of learning the 3D flow motion from double-frame particle sets in PTV experiments.

Deep learning in fluid mechanics has attracted increasing attention and developed rapidly over the past few years \cite{brunton2020machine}.
In the community of flow measurement, PIV has thrived due to the development of deep neural networks, especially  convolutional neural networks (CNN) \cite{cai2019dense, cai2019particle, liang2020filtering, lagemann2021deep}.
However, due to the difficulty of applying CNN-based approaches to cope with the unstructured grid particle data, PTV has until recently benefited few from deep learning and conversely still remains room for improvement. 
A recent work on learning-based PTV has been proposed by \cite{mallery2020dense}, where a recurrent neural network is applied to deal with the particle linking issue. However, multiple frames of particle trajectories are required as input and a pre-trained model is specialized for one flow case. On the other hand, pioneered by the recent PointNet-based backbone \cite{qi2017pointnet}, numerous approaches have been put forwards to estimate scene flow (mostly rigid motion) from unstructured 3D point clouds \cite{liu2019flownet3d, puy2020flot, wu2020pointpwc, wei2021pv}.
Inspired by the scene flow learning network, a preliminary investigation has been conducted to estimate 2D flow in PTV analysis, which shows a comparable performance with classical state-of-the-art PTV approaches on accuracy, robustness and efficiency \cite{liang2021deepptv}.
In this work, we present how to extract spatial geometric information from the basic PTV data (merely 3D particle positions of two frames) with graph neural network (GNN) and generalize the scene flow  to complex non-rigid flow motion (more common in physical phenomena) using our proposed optimal-transport guided RNN framework. The designed deep learning model is called GotFlow3D. 
The learnt flow by GotFlow3D can directly indicates the physical dynamics or be implemented as a-priori knowledge of the flow field into predictor-corrector schemes \cite{dabiri2020particle} to boost the particle tracking.
Our propose method is a new paradigm in PTV technology to recover the complex particle motion, which plays a significant role for further dynamic behavior analysis in physical and biological systems.

%

\begin{figure*}[!th]
	\centering{\includegraphics[width=6.4in]{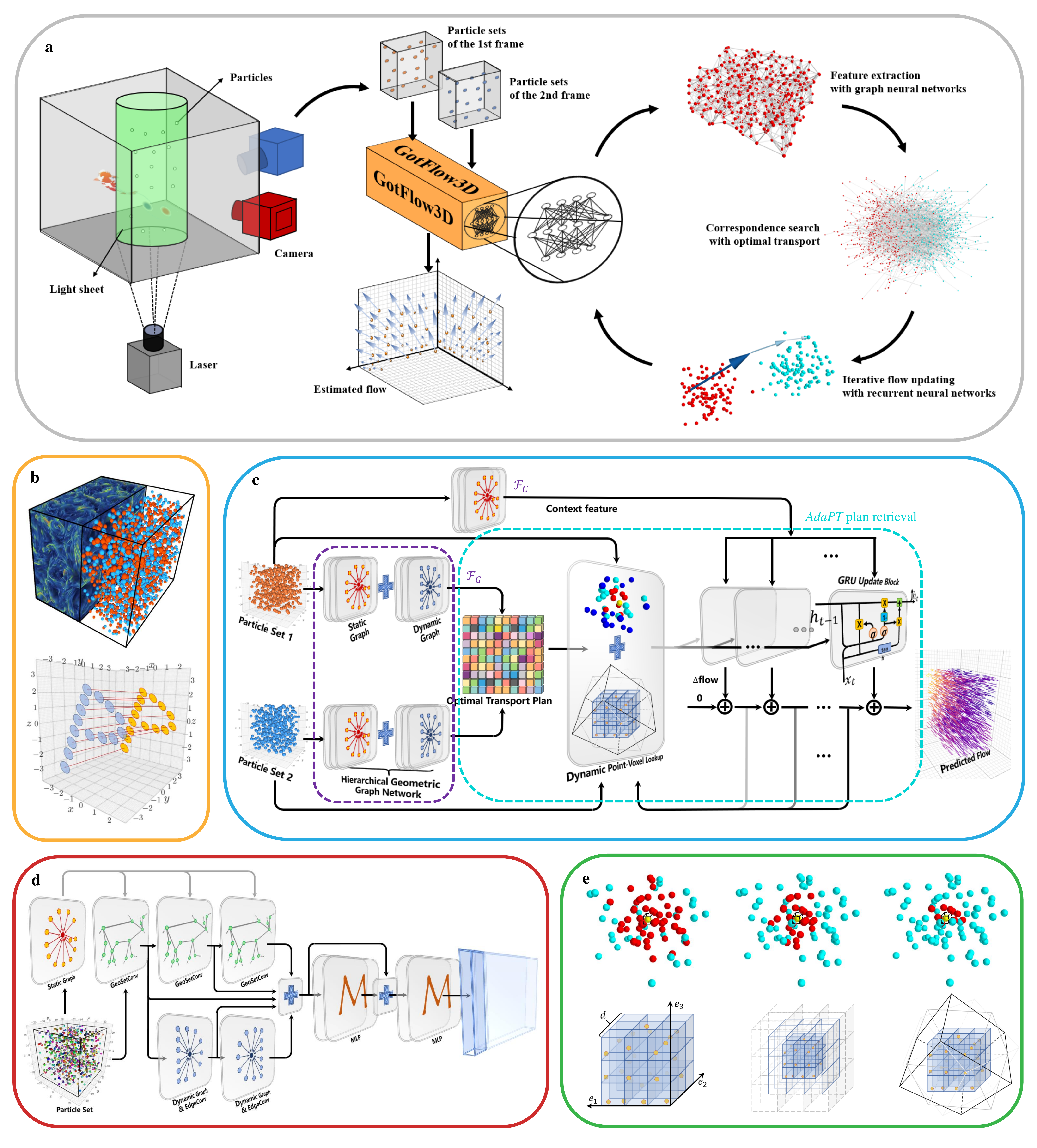}}
	\caption{\textbf{Overview of GotFlow3D for learning flow motion from particle sets. }  (a) Setup and the flowchart of GotFlow3D in particle tracking experiments. Double-frame particle distributions are obtained by tomographic configuration and then processed by GotFlow3D to output a 3D vector field.  (b) Illustration of the synthetic dataset generation. (c) Architecture of the proposed network, GotFlow3D, which consists of 3 key components: (1) graph-based feature extractor, (2) optimal transport, (3) recurrent GRU. (d) Visualization of \textit{SDGNN}, marked in purple box in (c), which fuses the static spatial graph and the dynamic feature graph to extract the geometric information of particle sets. (e) Illustration of the adaptive point-voxel optimal transport (\textit{AdaPT}) plan retrieval in the correspondence searching. The first raw of (e) depicts the searching of dynamically reduced nearest neighbors (red balls) of the investigated point (the yellow ball) in the point-based transport plan retrieval, while the bottom one represents the adaptively-deformable multi-scale cuboids in the voxel-based retrieval.
		\label{fig:figure_method}}
\end{figure*}

\section{Methods}
\subsection{Problem Setup}
\label{sec:setup}

To better understand the principles of the proposed model, we first describe the formulation of the complex flow learning problem.
The whole framework is shown in Figure~\ref{fig:figure_method}(a).
The proposed model consumes, as inputs, two unorganized consecutive 3D particle sets $\mathcal{P}$, $\mathcal{Q}$ and produces the dense vector field $ \mathcal{V} $.
The particle sets can be presented as the source set $ \mathcal{P} = \{ p_{i} \in \mathbb{R}^{D_{in}} \}_{i=1}^{n_{1}} $ with $n_{1}$ particles and the target set $ \mathcal{Q} = \{ q_{j} \in \mathbb{R}^{D_{in}} \}_{j=1}^{n_{2}} $ with $n_{2}$ particles, where each particle $p_{i}$ (or $q_{j}$) suspends in various flow media and faithfully follows the media flow motion.
Note that the attributes of each particle $p_{i}$ (or $q_{j}$) may contain the 3D coordinate $x_i \in \mathbb{R}^3$ together with some additional state features $f_i \in \mathbb{R}^c$ (e.g., intensity or color information).
To enhance the generalization for various flow measurement tasks, we only consider the 3D coordinate (i.e., $D_{in}=3 $) in this work. 

Each element $v_i $ in $ \mathcal{V} = \{ v_{i} \in \mathbb{R}^3 \}_{i=1}^{n_{1}}$ represents the predicted flow motion corresponding to each particle $p_i$.
Given the source set $ \mathcal{P}$ and the target set $ \mathcal{Q}$, the iteratively updated flow $ \mathcal{V}_t $ (in the $t$-th iteration) gradually translates the source $ \mathcal{P}$ to $\mathcal{P}'_t$, which is expected to approach the target $ \mathcal{Q}$.
Before the iterative process, we first employ the GNNs to extract the geometric pointwise feature $\mathcal{F}_{\mathcal{G}}$, which is adopted to assign the optimal transport plan $\mathcal{T} $.
Then, the recurrent neural network updates the flow by adaptively seeking the neighbors of $\mathcal{P}'_t$ in $ \mathcal{Q}$ and following the guidance of the corresponding optimal transport plan.

To promote the learning of complex non-rigid fluid flow beyond scene flow, the properties of the flow motion in fluids should be taken into consideration.
For examples, multi-scale eddies and chaotic property changes frequently exist in the complex fluid flow, which poses a great challenge to flow estimators.
The challenge encourages an emphasis on both the extraction of large-scale motion fields (e.g., laminar flow) and the refinement of small-scale structures (e.g., vortex).
In addition, the spatial local structures assembled with neighboring particles show approximately motion invariance and geometric similarities due to the local smooth fluid motion. All these properties are considered while designing the deep neural network. 


\subsection{GotFlow3D}

This section introduces the proposed method used to learn the complex flow motion from two consecutive
3D particle sets.
As shown in Figure \ref{fig:figure_method}(c), this model is designed to extract the geometric patterns of the 3D particle sets in the flow media,
make the optimal transport plan that encodes the point correspondences, and iteratively undate the flow motion by adaptively retrieving the corresponding transport plan between the two particle sets.
In the following, we present the architecture of GotFlow3D, which is constructed by three essential modules performing 
(1) graph-based feature extraction; 
(2) adaptive optimal transport plan lookup;
and (3) iterative flow updating. 
Eventually, we also describe the implementation and training details of GotFlow3D.

\subsubsection{Graph-based Feature Extraction}
To ensure an accurate flow estimation, it is of great importance to capture reliable pointwise features, which are supposed to be discriminative enough for exact correspondences searching.
Considering that the 3D coordinate is the only available input, the key point is to learn pointwise features representing the local geometry information.
To achieve the effective feature exploration on the unstructured and irregular input format, we construct two kinds of geometric graphs on the particle sets and employ the static-dynamic fusion GNN (\textit{SDGNN}) to derive the deep features of each particle.
Specifically, we construct the static graph $\mathcal{G}^{s}$ on 3D spatial space and the dynamic graph $\mathcal{G}^{d}$ on high-dimensional feature space.

We first define the static graph $\mathcal{G}^{s} = (\mathcal{V}_{er},\mathcal{E}^s)$, where the vertices  $\mathcal{V}_{er}=\{1, 2, ..., n_1(n_2)\}$ are composed of the entire given particle set $ \mathcal{P}$ (or $ \mathcal{Q}$).
The edges $\mathcal{E}^s \subseteq \lvert \mathcal{V}_{er} \rvert \times \lvert \mathcal{V}_{er} \rvert $ are yielded by connecting each particle (vertex) $p_i$ to its $k$-nearest spatial neighbors $\forall p_{i_k}\in \mathcal{N}_k^s(p_i)$ with the \textit{3D Euclidean distance} metric. 
The edges combining $p_i$ and its neighbors form local geometric structures and further generate the associated vertex feature $\tilde{f}_i =[p_{i_k}-p_i;r_{i_k};\theta_{i_k};\varphi_{i_k}] $, $\tilde{f}_i \in \mathbb{R}^{k\times6} $, where $p_{i_k}-p_i$ denotes the 3D directional vector.
In addition, $r_{i_k}, \theta_{i_k}$ and $\varphi_{i_k} $ represent the radial distance, azimuthal angle and polar angle in spherical coordinates, respectively, which supplement the relative spatial relationship of the neighboring vertices from another perspective.
Such spherical representations further provide rotation invariant correspondences, which are advantageous in  the estimation of  complex motions and larger deformations \cite{yang2022serialtrack}. 
The static graph  for a particle set needs to be computed only once before training and is kept for the following retrieval.

As shown in Figure \ref{fig:figure_method}(d), given $\mathcal{G}^{s}$ and $\mathcal{P}$, we employ the \textit{SDGNN} to encode the local geometric characteristics to generate the distinctive feature descriptor for each particle.
We denote the extracted pointwise feature of vertex $p_i$ at the $\beta$-th layer as $g_i^{\beta} \in \mathbb{R}^{D_{\beta}} $.
In the deep hierarchical \textit{SDGNN}, graph convolution operations are performed on the local geometric structure of each vertex $p_i$ to map the vertex feature $g_i^{\beta-1}$ of the previous layer to a new feature $g_i^{\beta}$ in the subsequent layer.
Inspired by the PointNet++ framework \cite{qi2017pointnet++}, we propose a \textit{GeoSetConv} layer to further embed the vertex features of the static graph to a high-dimensional feature space to better characterize the local geometric structures:
\begin{equation}\label{e1:gnn2}
\begin{aligned}
{^{s}g_i^{\beta}} =  \mathop{\max}_{j \in \mathcal{N}_k^s(p_i)} \ \left\{ \phi^{\beta} \left(\tilde{f}_i, \; {^{s}g_j^{\beta-1}}- {^{s}g_i^{\beta-1}} \right) \right\},
\end{aligned}
\end{equation}
here ${^{s}g_i^{\beta}}$ denotes the feature extracted from the static graph at the $\beta$-th layer, and $\mathcal{N}_s^k(p_i) $ depicts the $k$-nearest neighbors of vertex $p_i$ in the static graph.
In addition, $\phi^{\beta}$ is the $\beta$-th \textit{GeoSetConv} layer, which is implemented as multi-layer perceptrons (MLPs).
And $\max$ denotes the max-pooling operation, which aggregates features over the $k$ neighbors.
The $^{s}g_i^0$ is initialized by the 3D coordinate of $p_i$ in the first \textit{GeoSetConv} layer.

To further enrich the feature representations describing the detailed topological information, we explore the feature learning on the dynamic feature graph, as \textit{EdgeConv} did in \cite{wang2019dynamic}.
It is the first attempt at applying the dynamic graph CNN to extract features in the flow learning task.
The following experiments also demonstrate that the introduction of the dynamic feature graph contributes to the improvement of the accuracy.
The dynamic graph $\mathcal{G}^{d}$ is constructed and updated by aggregating the $k$-nearest neighbors on the high-dimensional feature space generated at each layer.
Specifically, for each layer, we build the dynamic graph $\mathcal{G}^{d}= (\mathcal{V}_{er},\mathcal{E}^d)$, where $\mathcal{G}^{d}$ shares the same vertices as that of $\mathcal{G}^{s}$ and the edges $\mathcal{E}^d$ link feature level neighbors to form local structures in the feature manifold.
Then, we formulate the update of the vertex feature ${^{d}g_i^{\beta}}$ of $p_i$ on the dynamic graph at the $\beta$-th layer as
\begin{equation}\label{e1:gnn3}
\begin{aligned}
{^{d}g_i^{\beta}} =  \mathop{\max}_{j \in \mathcal{N}_k^f(p_i)} \ \left\{ \varphi^{\beta} \left({^{d}g_j^{\beta-1}}, \; {^{d}g_j^{\beta-1}}- {^{d}g_i^{\beta-1}}\right) \right\},
\end{aligned}
\end{equation}
here $\mathcal{N}_k^f(p_i) $ depicts the index of the $k$-nearest neighbors of $p_i$ in the feature space.
And  $\varphi^{\beta}$ denotes the $\beta$-th \textit{EdgeConv} layer, which is also implemented as MLPs.
Different from the static graph, whose receptive field is limited in the local neighborhood and hinders the acquirement of long-term dependencies, the dynamic graph enlarges the receptive field to the whole input set.
The dynamic graph learns to construct a special receptive field by assigning more attention to the neighbors with similar feature (flow) properties.

Then, the proposed hierarchical \textit{SDGNN} incorporates the extrinsic geometric topology and the intrinsic feature correspondences by fusing the learnt features from the two types of graphs.
More precisely, we further concatenate the hierarchical features of these graphs from different layers of the GNN.
Finally, several MLPs with a skip connection are employed to derive the final deep pointwise feature 
\begin{equation}\label{e1:gnn1}
\begin{aligned}
\mathcal{F}_{\mathcal{G}} =  {\psi} \left(\mathop{\oplus}_{\beta} \left({^{s}g^{\beta}} , {^{d}g^{\beta}} \right) \right) ,
\end{aligned}
\end{equation}
where ${\psi}$ depicts the MLPs and $\mathop{\oplus}\limits_{\beta} (\cdot,\cdot)$ denotes the concatenation of features from different graphs and different layers.
The extracted pointwise feature is adopted for the following optimal transport planning to infer the particle correspondences via the feature similarity.
Furthermore, we employ another feature encoder to extract the context feature $\mathcal{F}_{\mathcal{C}}$ of $\mathcal{P}$, which provides additional context information to the flow update.
Such context feature encoder simply consists of several \textit{SetConv} layers from PointNet++, which only operates on the spatial static graph.

\subsubsection{Adaptive Point-Voxel Optimal Transport}
\textbf{Optimal transport. } 
Optimal transport theory \cite{villani2009optimal} guides the transportation and allocation between two distributions.
In this work, we regard the correspondences searching of two particle sets as the transportation and allocation of the particle mass, where each particle $p_i$ ($q_j$) is set to have a mass of $n_1^{-1}$ ($n_2^{-1}$).
Then, the cost has to be paid to transport each unit of mass from $p_i$ to $q_j$.
Therefore, the key issue of optimal transport is to seek a transport plan transforming a source particle distribution $\mathcal{P}$ to a target particle distribution $\mathcal{Q}$ whilst minimizing the transport cost \cite{peyre2019computational}.
In more detail, the computation of a transport plan $\mathcal{T} \in \mathbb{R}_{+}^{n_1\times n_2}$ between two particle sets can be formulated as 
\begin{equation}\label{e2:ot1}
\begin{aligned}
&\mathcal{T} = \mathop{\arg\min}\limits_{U \in \mathbb{R}_{+}^{n\times n}} \sum_{i,j=1} C_{ij} U_{ij} \\
&s.t. \quad U \textbf{1}_{n_1} = \bm{\mu}_p, \ U^T \textbf{1}_{n_2} = \bm{\mu}_q,
\end{aligned}
\end{equation}
where $C_{ij}$ is the transport cost from particle $p_i \in \mathcal{P}$ to $q_j \in \mathcal{Q}$, and $U_{ij}$ denotes the assigned quantity of mass to be transported from $p_i$ to $q_j$.
In addition, we define $\bm{\mu}_p = \frac{1}{n_1} \textbf{1}_{n_1}$ and $\bm{\mu}_q  = \frac{1}{n_2} \textbf{1}_{n_2}$ in the equality constraints.
The constraint manifests that the mass of each particle $p_i$ from the source distribution is supposed to be completely transported to the target and each particle $q_j$ of the target distribution is expected to be filled up with the transported mass.

However, due to the measurement noise in real experiments such as 3D reconstruction errors and object occlusions, the map reflecting the correspondences between two particle sets is neither injective nor surjective.
Therefore, the mass constraints cannot hold ideally.
To cope with this problem, it is preferred to apply the constraint relaxation to solve the optimal transport problem.
We follow the formulation proposed by FLOT \cite{puy2020flot} to employ the Sinkhorn algorithm \cite{cuturi2013sinkhorn} to efficiently optimize the transport plan with an entropic regularization and a mass regularization:
\begin{equation}\label{e2:ot2}
\begin{aligned}
\mathcal{T} = \mathop{\arg\min}\limits_{U \in \mathbb{R}_{+}^{n\times n}} 
&\left[\sum_{i,j=1}^N C_{ij}U_{ij} + \epsilon U_{ij}(\log U_{ij}-1) \right] \\
& + \lambda KL(U \textbf{1}_n, \bm{\mu}_p) + \lambda KL(U^T \textbf{1}_n, \bm{\mu}_q),
\end{aligned}
\end{equation}
here $KL$ represents the $KL$-divergence. 
In addition, $\epsilon$ and $\lambda $ denote the entropic and the mass regularization parameter, which are also learnt through the network training.
A small $\epsilon$ leads to a sparse transport plan, while a high $\lambda $ tends to reinforce the effect of the mass constraint.
Then, we define the transport cost between all pairs as the dissimilarity of the extracted pointwise feature via the cosine similarity:
\begin{equation}\label{e2:ot3}
\begin{aligned}
C_{ij} = 1 - \frac{\mathcal{F}_{\mathcal{G}}(p_i)^T \mathcal{F}_{\mathcal{G}}(q_j)}{\Vert \mathcal{F}_{\mathcal{G}}(p_i)\Vert_2 \Vert \mathcal{F}_{\mathcal{G}}(q_j)\Vert_2},
\end{aligned}
\end{equation}
where $\Vert \cdot \Vert_2$ denotes the $l_2$-norm.
A small cost between two particles generally corresponds to a high plan, which implies a latent possibility of particle correspondences.
The transport plan is constructed only once and is kept available for the following retrieval, which conduces to avoid numerous repeat matrix calculations.

\textbf{Adaptive  point-voxel optimal transport. } 
In this paper, we propose a novel approach to adaptively retrieve the optimal transport plan for the iterative flow updating.
For each iteration $t $, given the currently estimated flow $ \mathcal{V}_t $, the source $ \mathcal{P}$ is translated to $\mathcal{P}'_t$, where $P'_t = P+\mathcal{V}_t$.
The estimated flow is initialized at $ \mathcal{V}_0 =0$ and accumulated by the previous flow updates as
\begin{equation}\label{e4:gru1}
\mathcal{V}_t = \sum_{l=1}^{t}\Delta f_l,
\end{equation}
where $l$ denotes the index of the iteration.
Next, we can consult the transport plan between $\mathcal{P}'_t$ and $\mathcal{Q}$ to guide the flow update $\Delta f_{t+1}$ of the next iteration.
More specifically, the consultation is adopted by searching the neighbors of the warped source $\mathcal{P}'_t$ in the target $\mathcal{Q}$ and retrieving the corresponding transport plan $\mathcal{T}_t$. 
This iterative transport plan consultation contributes to lowering the difficulty of long range motion estimation and promote the refinement of small scale motion.

An essential step in the above-mentioned algorithm is: how to efficiently retrieve abundance and reliable information ($\mathcal{T}_t$) from the full transport plan ($\mathcal{T}$) through the neighborhood relation reasoning of the unstructured particle set.
%
To address this issue, the proposed approach further generates the adaptive point-voxel optimal transport (\textit{AdaPT}) plan $\mathcal{T}^*_t$ during each consultation based on the original point-voxel operation \cite{wei2021pv}.
More precisely, we integrate the captured point-based transport plan $\mathcal{T}^*_p$ and the voxel-based transport plan $\mathcal{T}^*_v$ to offer the long-range and small-scale correspondence of particle pairs in the consultation:
\begin{equation}\label{e3:pv1}
\begin{aligned}
\mathcal{T}^*_t = \mathcal{T}^*_p(P'_t,Q) + \mathcal{T}^*_v(P'_t,Q).
\end{aligned}
\end{equation}
First, we present the generation of the point-based transport plan $\mathcal{T}^*_p$ using a dynamic $k_t$-nearest neighbors searching, where the number of the located neighbors are gradually reduced as the iteration progressed.
Such construction adaptively narrows the search region to refine the fine-grained correspondence along with the increasing of iterations.
From the deep point of view, the learning of the subsequent smaller residual flow amounts to cope with particles with relatively larger spatial distance.
Accordingly, we dynamically reduce the searched neighboring particles to adapt to sparsely seeded regions, in which small number of neighbors perform better as noted in \cite{yang2022serialtrack}.
The computation of $\mathcal{T}^*_p$ and the attenuation of $k_t$ are formulated as:
\begin{equation}\label{e3:pv3}
\begin{aligned}
\mathcal{T}^*_p = \mathop{\max}_{j \in \mathcal{N}_Q^{k_t}(P'_{t})} \Phi_1 (\mathcal{T}(Q_j), \ Q_j - P'_t),
\end{aligned}
\end{equation}
\begin{equation}\label{e6:pv6}
k_t = k_0 - \eta  t,
\end{equation}
here $\mathcal{N}_Q^{k_t}(P'_t)$ denotes the index set of the $k_t$-nearest neighbors of $P'_t$ in $Q$, and $\mathcal{T}(Q_j)$ is the corresponding retrieved transport plan between $Q_j$ and $P'_t$.
In addition, we further combine the retrieved $\mathcal{T}(Q_j)$ and the relative spatial relation $Q_j - P'_t$ of the corresponding neighbors to deep encode the final plan $\mathcal{T}^*_p$, where the local spatial relation plays a similar role as the geometric attention weight.
The deep encoding is performed by the MLPs $\Phi_1$ with a $\max$ operation over the $k_t$ dimension.
Furthermore, $k_0$ is the initial value of $k_t$, while $\eta$ is set to a constant value.

To further cope with the large-scale motion embedded in the long-range correspondence of transport plan, we innovatively construct adaptively deformable multi-scale cuboids, which compartmentalize the geometric space containing particles into pyramid voxels to yield the voxel-based transport plan $\mathcal{T}^*_v$.
Specifically, each constructed cuboid is centered at each particle of $P'_t$ and collects the enveloped neighbors in $Q$.
As depicted in Figure~\ref{fig:figure_method}(e), the cuboid is composed of $d^3$ identical sub-cuboids, whose length ($\vec e_1$), width ($\vec e_2$) and height ($\vec e_3$) vectors are adaptively defined according to the several recent flow updates as:
\begin{equation}\label{e3:pv5}
\begin{aligned}
\begin{cases} 
\vec e_1 = (\frac{1}{8} \mathcal{S}(\lvert\Delta f_l \rvert) + \frac{1}{8} \cos\left \langle \Delta f_l, \Delta f_{l-1} \right \rangle + 1)  r  \vec l_1 \vspace{1ex}, \\
\vec e_2 = (\frac{1}{8} \mathcal{S}(\lvert\Delta f_{l-1}\rvert) + 0.9)  r  \vec l_2 \vspace{1ex}, \\
\vec e_3 = 0.9  r  \vec l_3, 
\end{cases}
\end{aligned}
\end{equation}
where $\mathcal{S}(\lvert\Delta f_l \rvert)$ and $\mathcal{S}(\lvert\Delta f_{l-1}\rvert)$ denote the normalized length of the current updated flow and the last updated flow over all particles in $P'_t$, respectively.
Such normalization encourages a longer-range correspondence search when large variations appear on the flow.
And $cos\left \langle \Delta f_l, \Delta f_{l-1} \right \rangle$ calculates the cosine similarity between $\Delta f_l$ and $\Delta f_{l-1}$.
Such similarity calculation also reinforces a longer-range correspondence search when recent flow updates keep  consistent but prompts a smaller-range correspondence search when they remain inconsistent.
Moreover, $\vec l_1$, $\vec l_2$ and $\vec l_3$ are unit orthogonal basis vectors of the three principal orientations used to guide the search direction of the correspondence, which are established as:
\begin{equation}\label{e7:pv7}
\begin{aligned}
\begin{cases} 
\vec l_1 = \Vert \Delta f_l \Vert_n \vspace{1ex}, \\
\vec l_2 = \Vert \Delta f_{l-1} - \frac{\left \langle \Delta f_{l-1} , \Delta f_{l} \right \rangle}{\left \langle \Delta f_{l} , \Delta f_{l}  \right \rangle} \Delta f_{l} \Vert_n \vspace{1ex}, \\
\vec l_3 = \Vert \vec l_1 \otimes \vec l_2 \Vert_n, 
\end{cases}
\end{aligned}
\end{equation}
where $\left \langle \cdot , \cdot \right \rangle$ denotes the inner  product of two vectors, while $\otimes $ represents the outer  product; $\Vert \cdot \Vert_n $ means that the vector is normalized to unit length. 
Furthermore, we define a basic scale parameter $r$, which controls the size deformation of the cuboid:
\begin{equation}\label{e3:pv8}
\begin{aligned}
r = 2^{s-1}  d_{m}  \alpha,
\end{aligned}
\end{equation}
where $d_{m}$ denotes the mean distance between each particle and its $k_d$-nearest neighbors in the input set, which is utilized to obtain the local particle concentration information.
And $\alpha $ is a learnable scale parameter optimized with the network training, which contributes to an adaptive process of particle sets with various particle densities.
In addition, $s$ is the index of the pyramid iteration, which gradually doubles the length size of the sub-cuboid to retrieve the multi-scale voxel-based plan.
The adaptive variation of the multi-scale cuboid size $r$ makes it possible to cover farther desired particles.

As illustrated above, the constructed cuboid presents the powerful deformability to adapt to current flow updates and input particle densities, while showing puissant capabilities to capture long-range correspondences with the multi-scale strategy.
Then, the cuboid partitions the enveloped neighboring particles in $Q$ into $d^3$ sub-cuboids, which indicate the relative direction of the neighbours to the central particle in $P'_t$.
Furthermore, the neighboring particles and the corresponding transport plan in each sub-cuboid are aggregated and averaged for the final calculation of $\mathcal{T}^*_v$:
\begin{equation}\label{e3:pv4}
\begin{aligned}
\mathcal{T}^*_v = \Phi_2 \left( \mathop{\uplus}_s \left( \mathop{\uplus}_b \left( \overline{\mathcal{T}} \left(    
^s\mathcal{N}_Q^{b} \left( P'_{t} \right)   
\right) \right) \right) \right),
\end{aligned}
\end{equation}
here $^s\mathcal{N}_Q^{b} \left( P'_{t} \right)$ denotes the neighboring particles of $P'_{t}$ in $Q$ enveloped in the $b$-th sub-cuboid for the $s$-th pyramid iteration.
And $\overline{\mathcal{T}}$ is the average over the transport plan of particles in the $b$-th sub-cuboid.
Moreover, $\mathop{\uplus}_s$ and $\mathop{\uplus}_b$ denote the concatenation over all pyramid iterations and all sub-cuboids, respectively.
There followed another MLPs $\Phi_2$, which is used to implement the similar deep encoding for the final plan $\mathcal{T}^*_v$.

\textbf{GRU flow refinement. }
Inspired by RAFT~\cite{teed2020raft}, we finally employ a recurrent GRU to estimate the flow update $\Delta f_{t+1}$, which mimics the iterative strategy in the optimization algorithm.
The applied GRU consumes the integrated \textit{AdaPT} plan $\mathcal{T}^*_t$, the current flow estimation $\mathcal{V}_{t}$, the context feature $\mathcal{F}_{\mathcal{C}}$ and the hidden state $h_{t-1}$ transmitted from the last iteration, while yielding a new hidden state $h_t$ encoding the future flow information:
\begin{equation}\label{e4:gru1}
z_t = \sigma (Conv_{1d} ([h_{t-1}, x_t], W_z)),
\end{equation}
\begin{equation}\label{e4:gru2}
r_t = \sigma (Conv_{1d} ([h_{t-1}, x_t], W_r)),
\end{equation}
\begin{equation}\label{e4:gru3}
\tilde{h_t} = \tanh (Conv_{1d} ([r_t \odot  h_{t-1}, x_t], W_h)),
\end{equation}
\begin{equation}\label{e4:gru4}
h_t = (1-z_t) \odot h_{t-1} + z_t \odot \tilde{h_t},
\end{equation}
where $x_t$ denotes the concatenated feature of $\mathcal{T}^*_t$, $\mathcal{V}_{t}$ and $\mathcal{F}_{\mathcal{C}}$.
And $\odot$ is the element-wise product.
In addition, $W_{z(r,h)}$ denotes the weights of convolutional layers in different steps.
Then, the $h_t$ is passed through a \textit{SetConv} layer and several convolutional layers to predict the flow update $\Delta f_{t+1}$.

\subsubsection{Implementation Details}

In this framework, there are some predefined parameters in the graph construction and the \textit{AdaPT} plan retrieval process.
The number of the searched neighboring particles used to construct the static graph and the dynamic graph is set as $k=32$.
For the \textit{AdaPT} plan retrieval, the dynamic $k_t$ is initialized as $k_0=32$, while $\eta$ is fixed to 2.
And when the iteration exceeds 8, the $k_t$ is fixed to 16.
The resolution of the cuboid $d$ and the number of the pyramid iteration $s$ are both set as 3.
In addition, the $k_d$ is set as 3 to calculate the mean local particle distance.
Moreover, the scale parameter $\alpha$ is initialized as 0.5 and has a special learning rate, which is 0.1 times the global learning rate, to enhance the training stability.
The dimension of the final pointwise feature extracted by the hierarchical GNN is set as 128.
Furthermore, the $\epsilon$ and $\lambda$ used to control the two regularizations are initialized as 0 and 1, respectively. 
And the flow is iteratively updated for 8 iterations.

The network training are performed with supervised loss using the entire estimated flow sequence $\{\mathcal{V}^e_1, ... \mathcal{V}^e_T \}$:
\begin{equation}\label{e5:loss}
\mathcal{L} = \frac{1}{N} \sum_{i=1}^N \sum_{t=1}^T \gamma_t \Vert \mathcal{V}^e_t - \mathcal{V}^{gt} \Vert,
\end{equation}
where $\mathcal{V}^e_t$ denotes the flow estimation of the $t$-th iteration, while $\mathcal{V}^{gt}$ is the ground truth flow.
In addition, $\Vert \cdot \Vert$ denotes the $l_1$ distance, and $T$ is the total number of flow update iterations. 
Moreover, $N$ denotes the total number of the particles. Mini-batch training is employed with a batch size of 16.
Then, we further impose an exponentially increased weight $\gamma_t$ to the estimated flow sequence:
\begin{equation}\label{e5:loss}
\gamma_t = \mu ^{T-t},
\end{equation}
where $\mu$ is set to 0.8.
The network is trained on six TITIAN XP GPUs for 40 epoches using Adam optimizer \cite{kingma2014adam} with a initial learning rate of 0.001.

\subsection{Datasets}

\subsubsection{Synthetic training datasets }
\label{sec:syntheticdataset}

Training data with ground-truth information plays an indispensable role in the network optimization with supervised learning strategies.
However, the difficulty of recovering highly accurate ground-truth flow from real data enforces us turn to the simulated data from computational fluid dynamics (CFD), which is capable of generating various flow cases representing the true fluid mechanical characteristics by solving the Navier–Stokes equations \cite{raffel2018particle}.

To perform the flow supervision, we create the first synthetic 3D fluid flow dataset named \textit{FluidFlow3D}, which is characterized by two consecutive particle sets $\mathcal{P}$, $\mathcal{Q} $ and the corresponding ground-truth flow set $\mathcal{V}^{gt}$.
In particular, we first generate a particle set $\mathcal{P}$, which comprises more than two thousand randomly distributed particles in the 3D observation volume.
The observation volume is also defined in the random location of the CFD computational domain.
Next, we use the public CFD data from Johns Hopkins Turbulence Database (JHTDB) \cite{li2008public}, which is proven to provide physically correct simulated flow structures, to generate the ground-truth flow $\mathcal{V}^{gt}$.
This is achieved by applying the built-in functions to query the JHTDB database with the second order accurate Runge-Kutta integration scheme in time and the fourth order Lagrange interpolation in space, which are executed on the database nodes.
The flow structures queried from JHTDB includes the  incompressible isotropic turbulent flow, the incompressible magneto-hydrodynamic (MHD) turbulence, the fully developed turbulent channel flow and the transitional boundary layer flow.
Therefore, we can get the next frame $\mathcal{Q}_0$ of the given particle set $\mathcal{P}$ using the integrated trajectory with a designated time interval, subsequently  get the the corresponding ground-truth flow $\mathcal{V}^{gt} = \mathcal{Q}_0 - \mathcal{P}$.
The sketch of this generation procedure is illustrated in Figure \ref{fig:figure_method}(b).
In addition to the data from JHTDB, several flow benchmarks, including the uniform flow and the Beltrami flow \cite{ethier1994exact}, are also generated to increase the diversity of the training dataset.
Eventually, the synthetic training dataset consists of six categories of typical flow cases.

Furthermore, to enhance the generalization capabilities of the trained network for realistic experimental configurations, we augment the above flow cases with various flow conditions. 
The measurement performance in PTV experiments is highly limited by   large dynamic velocity ranges and high particle densities. These two factors can be characterized by the particle displacement ratio $\rho$, which has been widely used in the PTV community \cite{maas1993particle} and  can comprehensively depict the flow conditions:
\begin{equation}\label{e7:ratio}
\rho = \frac{d_{a}}{u_{\max} \Delta t},
\end{equation}
here $d_{a} $ and $u_{\max}\Delta t $ denote the average distance between particles in the whole set (e.g., $\mathcal{P}$) and the maximum displacement between two input sets, respectively. It is reported that the classical flow estimators can easily achieve satisfactory performance for flow cases with $\rho \gg 1$, while facing more challenges when $\rho$ gets smaller \cite{maas1993particle}. 
In our training dataset, we change the characteristic observation volume to vary $d_{a} $ (the number of particles remains constant) and alter the time interval $\Delta t $ to obtain different maximum displacements. 
As a result, this important parameter $\rho$ in the synthetic dataset ranges from $0.35$ to $6.58$, much beyond the normal range of real PTV experiments. 
Moreover, to mimic the real-world applications, we add some noises to create more realistic cases, which mainly arise from the physical motion of particles in and out of the field of view, the measurement noise and the occlusion.
To this end, we  shuffle the particles in $\mathcal{Q}_0$ to obtain the second set $\mathcal{Q}$, which directly disorders the correspondence.
Then, we randomly replace $n\%$ of the particles in $\mathcal{Q}$ with new random particles.
Such noises eliminates $n\%$ of the correspondence of $\mathcal{P}$, while introducing new isolated particles in $\mathcal{Q}$ with no correspondence.
Here we set $n$ with $1\%$ level of noise for this dataset.
Finally, we generate a synthetic dataset with 16K training samples and 1.6K validation samples.
A table summarizing the synthetic dataset is given in Table~\ref{tab1:dataset}. 

\subsubsection{Testing Dataset} 

A family of synthetic \textit{FluidFlow3D}-derived datasets, a public turbulent cylinder wake flow dataset (denoted by \textit{CylinderFlow}) \cite{khojasteh2022lagrangian} and a real-world dataset (denoted by \textit{SphericalIndentationFlow}) across different domains are adopted to conduct the experimental evaluations.

\textit{FluidFlow3D-family}. Apart from the aforementioned \textit{FluidFlow3D}, we further transform \textit{FluidFlow3D} to a new dataset with normalized spatial scale (denoted by \textit{FluidFlow3D-norm}) to compute the unified evaluation metric among flow cases with different observation volumes.
The \textit{FluidFlow3D-norm} is used for the comparison with the state-of-the-art methods and ablation studies.
To analyze the robustness, we create a synthetic subset \textit{FluidFlow3D-noise} with the different noise level.
Furthermore, another subset \textit{FluidFlow3D-ratio} regarding various $\rho$ is also generated.

\textit{CylinderFlow}. This public dataset contains Eulerian velocities and Lagrangian particle trajectories of the cylinder wake flow with a Reynolds number of 3900 \cite{khojasteh2022lagrangian}. 
Such flow data is also calculated using CFD.
In addition, the fourth-order Runge-Kutta scheme in time and trilinear interpolations in space are utilized to transport about 200,000 particles to obtain the Lagrangian trajectories.
Trajectories of two 3D observation volumes are available for the assessment of tracking algorithms.

\textit{DeformationFlow}. This experiment is conducted by measuring the volumetric deformation of a soft polyacrylamide (PA) hydrogel exerted by the spherical indentation \cite{yang2022serialtrack}. 
A stainless steel sphere is placed on the surface of the PA hydrogel, which leads to the indentation deformation due to the force of gravity.
Then the 3D volumetric image of the hydrogel material as well as the embedded fluorescent particles are scanned both before and after the deformation with multiphoton microscopy.

\subsection{Evaluation Metrics}

To quantitatively assess and compare the performance of all the methods, we adopt the widely used four evaluation metrics in scene flow learning tasks \cite{liu2019flownet3d, wu2020pointpwc, puy2020flot, wei2021pv}, including EPE, Acc Strict, Acc Relax and Outliers.
The end-point-error EPE averaged over each particle is defined as:
\begin{equation}\label{e5:loss}
EPE = \frac{1}{N} \sum_{i=1}^N  \Vert \mathcal{V}^e - \mathcal{V}^{gt} \Vert_2,
\end{equation}
where $\Vert \cdot \Vert_2$ denotes the $l_2$ distance.
The Acc Strict is calculated as the percentage of particles with  the relative EPE error $<$ 5\% , while the Acc Relax is defined as the percentage of particles with the relative EPE error $<$ 10\% .
In addition, the Outliers are denoted as the percentage of particles with EPE $>$ 0.3 unit or the relative error $>$ 10\%.
Moreover, in the comparison with tracking algorithms, two metrics are employed as the assessment principles for the tracking accuracy and tracking reliability, including the yield rate ($E_y$), the reliability rate ($E_r$):
\begin{equation}\label{e8:metric}
E_y = \frac{N_c}{N_g}, \qquad 
E_r = \frac{N_c}{N_e},
\end{equation}
where $E_y$ denotes the ratio of correct matches (denoted by $N_c$) to all ground-truth matches ($N_g$), while $E_r$ is the proportion of correct matches to all extracted matches ($N_e$).

\section{Results}
\label{sec:results}

In this section, a great deal of evaluations on experiments are presented to comprehensively assess the performance of GotFlow3D in the complex flow learning task.
We compare GotFlow3D to the state-of-the-art methods and verify the robustness to the noise as well as various displacement ratios.
There followed the validation on a public fluid flow dataset, where the superiority of GotFlow3D to traditional PTV methods and the promotion of GotFlow3D to the tracking accuracy are investigated.
Finally, a PTV experiment is conducted to further investigate the generalization of GotFlow3D to real-world experimental data.

\subsection{Comparison with SOTA scene flow learners }
\label{sec:sota}

\renewcommand\arraystretch{1.2}
\begin{table}[t]  
	\centering
	\caption{Performance comparison between the state-of-the-art scene flow learning methods and GotFlow3D. The best results are marked in bold. \label{tab2:soat}}
	\resizebox{\columnwidth}{!}
	{\begin{tabular*}{21.4pc}{lllll}
			\noalign{{\hrule height 5pt}} 
			\textbf{Methods}        &\textbf{EPE} &\textbf{Acc Strict}  &\textbf{Acc Relax} & \textbf{Outliers} \\
			\noalign{{\color{black}\hrule height 1pt}}
			FlowNet3D \cite{liu2019flownet3d} &0.06229   &19.34\%   &38.23\% &61.77\% \\
			\noalign{{\color{white}\hrule height 1pt}}
			FLOT \cite{puy2020flot} &0.05865   &24.99\%   &45.59\% &54.41\% \\  
			\noalign{{\color{white}\hrule height 1pt}}
			PointPWC-Net \cite{wu2020pointpwc} &0.01723   &46.75\% &71.61\%  &28.39\% \\        
			\noalign{{\color{white}\hrule height 1pt}}
			PV-RAFT \cite{wei2021pv} &0.01651 &72.98\% &83.69\% &16.31\% \\  
			\noalign{{\color{white}\hrule height 1pt}}
			GotFlow3D     &\textbf{0.00487} &\textbf{93.15\%} &\textbf{96.38\%} &\textbf{3.62\%} \\  
			\noalign{{\color{black}\hrule height 1pt}}
	\end{tabular*}}{}
\end{table}

We first present quantitative results on the test set of \textit{FluidFlow3D-norm} by comparing our model with other four representative state-of-the-art scene flow learning methods, including the FlowNet3D \cite{liu2019flownet3d}, FLOT \cite{puy2020flot}, PointPWC-Net \cite{wu2020pointpwc} and PV-RAFT \cite{wei2021pv}.
All of these baseline methods are trained on the training set of \textit{FluidFlow3D-norm} with the training hyper-parameters reported in their articles.

The comparisons of the EPE, Acc Strict, Acc Relax and Outliers metrics of these methods are listed in Table~1.
As revealed in the table, our proposed model outperforms other baseline methods and achieves the state-of-the-art performance on all of the four metrics.
Specifically, GotFlow3D surpasses FlowNet3D and FLOT by more than an order of magnitude on the EPE,  significantly boosts the Acc Strict and Acc Relax and brings the Outliers to a extremely low level.

\begin{figure*}[!ht]
	\centering{\includegraphics[width=6.6in]{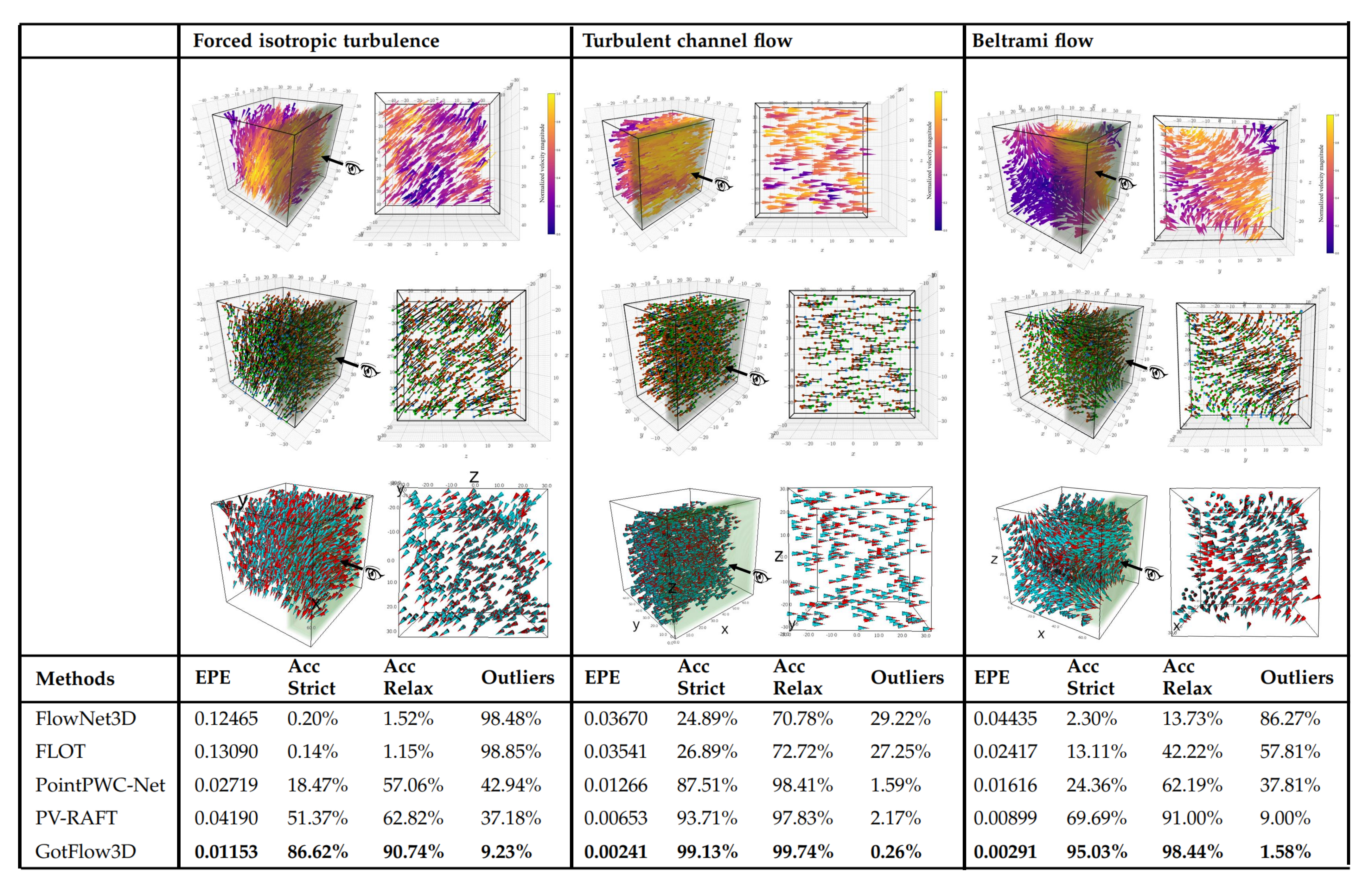}}
	\caption{\textbf{Flow estimation performance on testing examples in the \textit{FluidFlow3D} dataset. } Three flow cases are  demonstrated here, including the  Forced isotropic turbulence (left), Turbulent channel flow (middel) and Beltrami flow (right), while other cases can be found in Table~\ref{tab3:flowcase}. There are three kinds of visualization for each case: \textbf{(top)} the flow field estimated by GotFlow3D; \textbf{(middle)} the source particles $ \mathcal{P}$ (red), transported particles $ \mathcal{P'}$ (green) using the learnt flow and the corresponding target particles $ \mathcal{Q}$ (blue); \textbf{(bottom)} the ground-truth flow vectors (red) and the estimated flow vectors (cyan). In addition, the table shows the comparison of overall performances  between GotFlow3D and other SOTA scene flow learners on these testing cases. 
		\label{fig:synvis}}
\end{figure*}

To visualize more detials, we conduct comparative experiments over different flow cases in \textit{FluidFlow3D-norm} and verify the generalization and superiority of the proposed model. 
Three typical examples are demonstrated in Figure~\ref{fig:synvis}, while the errors for all the flow cases are given in Table~\ref{tab3:flowcase}. 
As shown in Figure~\ref{fig:synvis} and Table~\ref{tab3:flowcase}, GotFlow3D achieves the state-of-the-art performance with the smallest EPE for all flow cases. %
In addition, GotFlow3D presents the highest Acc Strict (Relax) and the smallest Outlier except for the Uniform flow case. 
In the Uniform flow case, we observe that PV-RAFT shows comparable results with GotFlow3D. PV-RAFT is slightly better, and they both achieve over 99\% accuracy and contain less than 0.02\% outliers. We suspect that such a good performance in this relatively simple case is attributed to the GRU module employed both in GotFlow3D and PV-RAFT. 
However, in some of the complex flow cases (e.g., Forced isotropic turbulence, Forced MHD turbulence and Beltrami flow) which contain abundant fine-grained small-scale flow structures, the other flow estimators including PV-RAFT provide worse results, 
whereas GotFlow3D consistently performs the best on the EPE and Outliers even coping with the difficult datasets containing complicated flow scenes and diverse motion patterns.

To further depict the learning performance of GotFlow3D, we visualize the estimated flow from GotFlow3D with diverse fashions in Figure~\ref{fig:synvis}.
Specifically, we plot the results of three test data samples from \textit{FluidFlow3D-norm}, including the Forced isotropic turbulence (left), Turbulent channel flow (middle) and  Beltrami flow (right).
Moreover, we present the predicted flow fields (top) to show the capability of GotFlow3D to learn complex flow distributions, which involve numerous small-scale flow structures.
In the middle row, the source particle $ \mathcal{P}$ (red), transported particles $ \mathcal{P'}$ (green) using the learnt flow and the corresponding target particle $ \mathcal{Q}$ (blue) are visualized to qualitatively demonstrate the tiny end point error, which also illustrates how GotFlow3D copes with the flow cases containing both small and large displacements. 
Furthermore, we also manifest both the ground-truth flow (red) and the flow estimations (cyan) in the bottom row.
It can be observed that the estimations maintain highly consistency with the ground truth.

\begin{figure*}[!th]
	\centering{\includegraphics[width=6.4in]{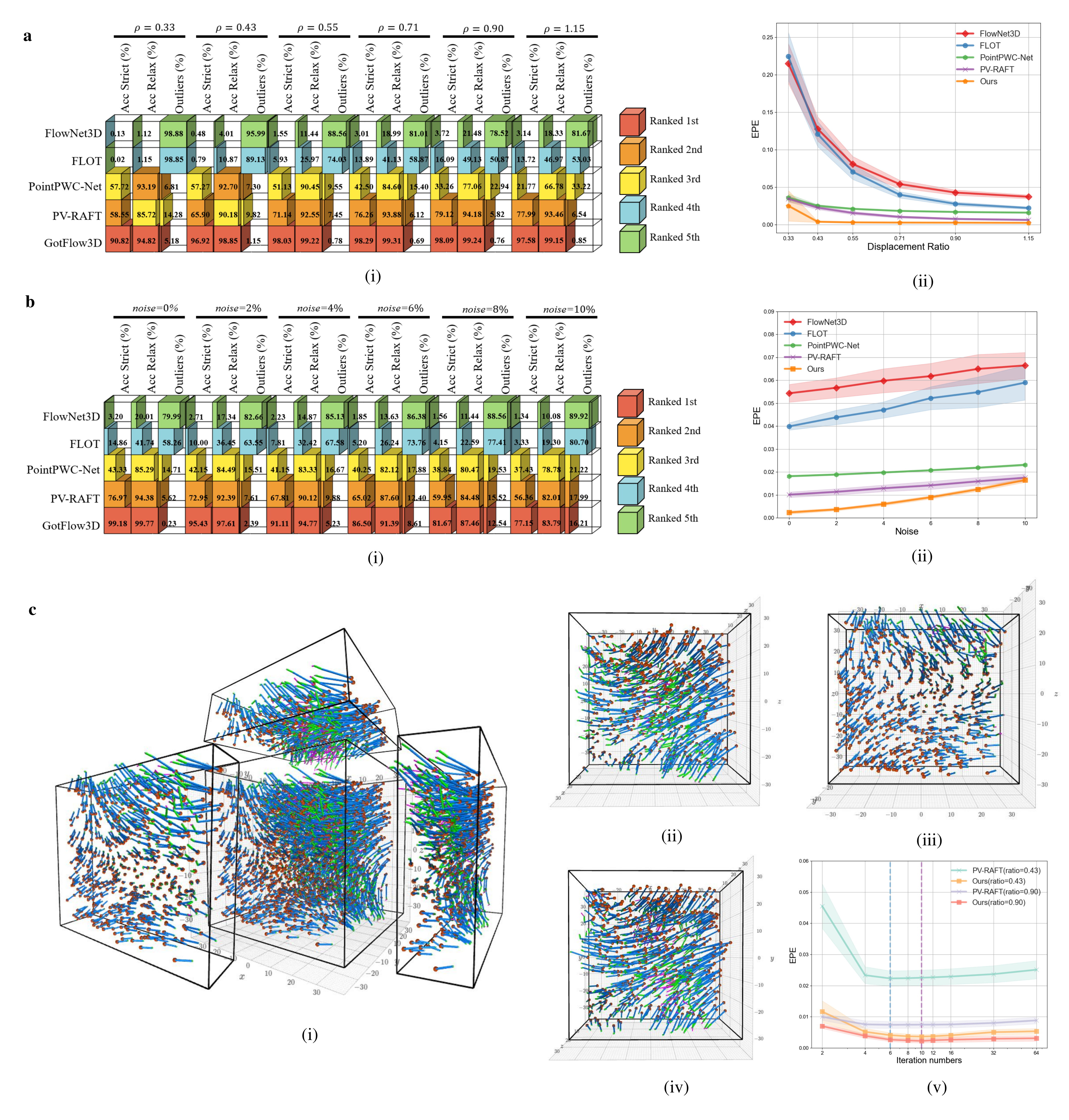}}
	\caption{\textbf{Robustness assessment for GotFlow3D. } (a) Performance of the  state-of-the-art scene flow learners with regard to the displacement ratio $\rho$. The right figure shows the mean (solid lines) and standard deviation (shadows) values, which are computed over 100 samples. (b) Performance regarding to various noise levels.  (c) Recurrent flow refinement estimated by GotFlow3D. Particles and motions  in different iterations are visualized: (red) source particles; (blue) flow motion estimated in iteration~1;  (green) updated residual flow from iteration 1 to 2;  (purple) updated residual flow from iteration 2 to 32.  Different viewing angles are demonstrated (i-iv). Moreover, the performance (EPE) of two recurrent learning models with respect to iteration numbers  are shown in (v). 
		\label{fig:robust}}
\end{figure*}

\subsection{Robustness against experimental conditions}

To further quantify the generalization and robustness of GotFlow3D facing extreme flow conditions, we conduct comparisons across different methods on dedicated datasets with wide range of  velocity  and varying  noise levels. 
The  metrics reported in this section are obtained by taking the average over 100 samples for each investigated condition.

To investigate the influence of the dynamic velocity range, we first present the assessment with respect to different displacement ratios $\rho$ on the subset \textit{FluidFlow3D-ratio}.
This data subset is generated by simulating the Beltrami flow with varied velocity magnitudes, which leads to a series of $\rho$ ranging from 0.33 to 1.15, covering most situations in real experiments. 
We demonstrate the Acc Strict, Acc Relax and Outliers of all flow learners in Figure~\ref{fig:robust}(a)(i), also  plot the corresponding EPEs as a function of $\rho$ in Figure~\ref{fig:robust}(a)(ii). 
As shown, GotFlow3D leads the way in EPE, Acc Strict (Relax) and Outliers compared to other methods while surpasses the FlowNet3D and FLOT to a great extent.
When facing the flow condition with a extreme low $\rho$ of 0.33, GotFlow3D consistently shows impressive performances. 
GotFlow3D presents a relatively higher variance on EPE than PV-RAFT and PointPWC-Net in the case when $\rho=0.33$ in Figure~\ref{fig:robust}(a)(ii), which may be due to that the $\rho$ of 0.33 is lower than the $\rho$ of the training set (from 0.35 to 6.58).
Other than that,  the EPE curve of GotFlow3D nearly remains flat with low variances when $\rho$ increases from 0.43 to 1.15.
We can observe that PV-RAFT also shows comparable results and approaches GotFlow3D when facing conditions with large displacement ratio (i.e., the case where flow estimators are easier to track particles). 
Nevertheless,  the outperformance of GotFlow3D convincingly reveals the robustness toward diverse flow cases with wide dynamic velocity ranges.

Likewise, the performance of all these flow estimators are demonstrated with respect to various noise levels defined in Section \ref{sec:syntheticdataset}.
This comparison is assessed on the subset \textit{FluidFlow3D-noise}, which is also generated using Beltrami flow with mean $\rho$ of 0.71. 
We investigate the noise level ranging from 0 \% to 10\%, which is reasonable to mimic the noises in real experiments. 
The evaluation metrics among different learning methods are shown in Figure~\ref{fig:robust}(b)(i), while the EPEs are ploted against the noise in Figure~\ref{fig:robust}(b)(ii). 
It is obviously seen that the metrics of all methods inevitably get worse with the increase of noise level, which is expected. 
Furthermore, it can be observed that GotFlow3D achieves the best performance on EPE, Acc Strict (Relax) and Outliers among all investigated noise levels and surpasses FlowNet3D as well as FLOT by a large margin (more than 50\%) on Acc Strict (Relax) and Outliers.
When it comes to a condition with high level noise, GotFlow3D does not possess much superiority over PV-RAFT while the gap between them gradually narrows. 
The impressive performance indicates the robustness of GotFlow3D facing some degree of noise levels.

To better illustrate the effects of the recurrent flow estimation, we investigate the performance of the network inference with different iteration numbers in the recurrent flow updating.
The iteration number is a user-defined parameter in inference stage, which can be set higher to learn flow with large displacements and can be defined lower to reduce the computational cost. 
This investigation is assessed on two flow subsets from \textit{FluidFlow3D-ratio}:  a more difficult case of $\rho = 0.43$ and an easier case of $\rho = 0.90$,  and the comparison is only performed between the learning methods with recurrent unit (i.e., PV-RAFT and GotFlow3D). 
As shown in Figure~\ref{fig:robust}(c)(v), both methods yield a remarkable improvement on EPE as the iteration number increases to 6.
However, the error of PV-RAFT becomes slightly higher  when the iteration number exceeds 6 (marked with a blue dashed line) for both values of $\rho$.
On the other hand, the optimal iteration number of GotFlow3D is 10 (marked with a purple dashed line) for both the flow cases.
These curves indicate that the iteration number has a profound impact on the performance of recurrent approaches. However, it can be found that the GotFlow3D is more accurate and less sensitive in different configurations. 
It also indicates that it is better to chose an iteration number close to the pre-defined value of the training process while applied to real flow scenarios. 
To understand the recurrent mechanism more intuitively, we visualize the recurrently updated flow  of GotFlow3D by plotting trajectories of the transported particles in several iterations. 
As shown in Figure~\ref{fig:robust}(c)(i-iv), the source particles (red) are transported using the estimated flow in iteration~1 (blue), the updated residual flow from iteration 1 to 2 (green) and the updated residual flow from iteration 2 to 32 (purple).
It can be observed that the major flow motion is estimated in the first two iterations while the residual flow continues to refined slowly in the remaining iterations.

In addition to the aforementioned assessments, an ablation study regarding different modules in GotFlow3D is also performed to demonstrate the effectiveness of each module. The results are summarized in Table~\ref{tab6:ablation}. 
We also compute the inference time of GotFlow3D to present the computational efficiency.
GotFlow3D takes about 0.1~s to estimate the flow of one sample containing about 2000 particles on a NVIDIA RTX 3090 GPU, which is promising for online flow visualization and measurement.

\subsection{GotFlow3D for shedding flow with a high Reynolds number}

PTV is  one of the most primary techniques to analyse 3D turbulent  flows by tracking particles. In this section, we perform further comparisons between GotFlow3D  and other advanced open-source tracking algorithms on a public dataset \textit{CylinderFlow}.
We should note that GotFlow3D is a motion estimator, which does not perform particle matching and linking as the open-source PTV algorithms do. 
Therefore,   we present the assessment by generalizing GotFlow3D as a customized motion predictor to provide existing PTV algorithms with additional motion prior information (i.e., initialization), which is supposed to promote the particle matching between adjacent frames.
Two open-source particle tracking algorithms, i.e., T-PT \cite{patel2018rapid} and SerialTrack \cite{yang2022serialtrack}, are implemented as evaluation benchmarks, which have been proven two of the state-of-the-art double-frame 3D tracking algorithms. 
T-PT is a topology-based particle tracking algorithm that encodes relative spatial neighbors as feature descriptors to deal with large displacements, while SerialTrack resolves large deformation and rotational motion with scale and rotation invariant augmented particle tracking. 
Both the two algorithms have been successfully applied to reveal complex motion behaviors in biological and physical applications \cite{leggett2020mechanophenotyping, scimone2018modular}.

\renewcommand\arraystretch{1.4}
\begin{table*}[t]
	\centering
	\caption{Performance comparison with other recently-developed particle  tracking algorithms for \textit{CylinderFlow} dataset.  The best results are marked in bold. The numbers (30, 60 and 90) in the first raw represent the time interval between two particle sets, and the corresponding displacement ratios are shown in the brackets. The terms ``GotFlow3D + X'' means that GotFlow3D is applied to provide initialization for the PTV algorithms. \label{tab7:ptv}} 
	\resizebox{\textwidth}{!}
	{\begin{tabular*}{50.65pc}{llllllllllllllll}
			\noalign{{\hrule height 5pt}} 
			\multirow{1}{*}{\makecell*[l]{ \quad }}  &\multicolumn{5}{l}{\textbf{30 ($\rho=0.1.06$)}}      &\multicolumn{5}{l}{\textbf{60 ($\rho=0.52$)}} & \multicolumn{5}{l}{\textbf{90 ($\rho=0.35$)}} \\
			\noalign{{\color{white}\hrule height 1pt}}
			\multirow{1}{*}{\makecell*[l]{ Methods }}  &\bm{$N_c$} &\bm{$N_g$} &\bm{$N_e$} &\bm{$E_y$}  &\bm{$E_r$} &\bm{$N_c$} &\bm{$N_g$} &\bm{$N_e$} &\bm{$E_y$}  &\bm{$E_r$} &\bm{$N_c$} &\bm{$N_g$} &\bm{$N_e$} &\bm{$E_y$}  &\bm{$E_r$} \\
			\noalign{{\color{black}\hrule height 1pt}} 
			T-PT \cite{patel2018rapid}   &161727 &195504 &161875 &82.72 &\textbf{99.91} &118865 &193903 &119046 &61.30 &\textbf{99.85} &98202 &192375 &98324 &51.05 &\textbf{99.88} \\
			\noalign{{\color{white}\hrule height 1pt}}
			GotFlow3D + T-PT &192196 &195504 &192450 &98.31 &99.86 &166232 &193903 &166908 &85.73 &99.60 &131451 &192375 &132087 &68.33 &99.52 \\
			\noalign{{\color{white}\hrule height 1pt}}
			SerialTrack \cite{yang2022serialtrack}  &181334 &195504 &189493 &92.75 &95.69 &123643 &193903 &187647 &63.76 &65.89 &95334 &192375 &\textbf{187501} &49.55 &50.84 \\ 
			\noalign{{\color{white}\hrule height 1pt}}
			GotFlow3D + SerialTrack  &\textbf{192350} &195504 &\textbf{194303} &\textbf{98.39} &98.99 &\textbf{178556} &193903 &\textbf{190297} &\textbf{92.08} &93.83 &\textbf{154322} &192375 &187211 &\textbf{80.21} &82.43 \\  
			\noalign{{\color{black}\hrule height 1pt}} 
	\end{tabular*}}{}
\end{table*}

To better show the promotion brought to these PTV algorithms from GotFlow3D, we conduct three comparisons with different frame intervals, including the first frame and the 30th frame (denoted as experiment 30), the first frame and the 60th frame (denoted as experiment 60), the first frame and the 90th frame (denoted as experiment 90). 
Larger time interval generally leads to relatively larger displacement motion, and the three investigated experiments  correspond to  $\rho $ of 1.06 (30), 0.52 (60) and 0.35 (90), respectively.
For evaluation of PTV algorithms, the yield rate ($E_y$) and reliability rate ($E_r$) are used as metrics for comparison, which strongly indicate the tracking accuracy and tracking reliability  in PTV applications.
As shown in Table~2, the $E_y$  and the correct matches $N_c$ of all methods decrease as the frame interval gets longer.
T-PT consistently presents the highest $E_r$ among all the three experiments, showing the tracking reliability of this method. However, the $E_y$ of T-PT is relatively low, which means it misses a large number of particle tracks. 
SerialTrack outperforms T-PT on $E_y$ and $N_c$ in conditions of high $\rho$ (experiment 30 and 60) but is slightly surpassed by T-PT in the challenging experiment 90.
Impressively, the integrated implementations named ``GotFlow3D+T-PT'' and ``GotFlow3D+SerialTrack'' improve the $E_y$ performance of T-PT and SerialTrack to an extremely high level, indicating that much more correct matches are extracted. 
In particular, the proposed GotFlow3D boosts the $E_y$ of T-PT by 15-24\% and that of SerialTrack by 6-30\%. 
The highest improvement of $E_y$ regarding T-PT comes from the experiment 60 (about 24\%), while that regarding SerialTrack is obtaiend from the experiment 90 (about 30\%). 
It can be seen that GotFlow3D yields larger promotion coping with the conditions with relatively larger displacements.
When facing the condition with short frame interval of 30 (i.e., high $\rho$ value), ``GotFlow3D+T-PT'' and ``GotFlow3D+SerialTrack'' achieve a nearly perfect performance with both $E_y$ and $E_r$ approaching 100\%.
The $E_r$ of ``GotFlow3D+T-PT'' and ``GotFlow3D+SerialTrack'' surpasses SerialTrack but is not superior to T-PT, since GotFlow3D also boosts the extraction of more matches, which may lead to the decrease of $E_r$.

\begin{figure*}[t]
	\centering{\includegraphics[width=6.8in]{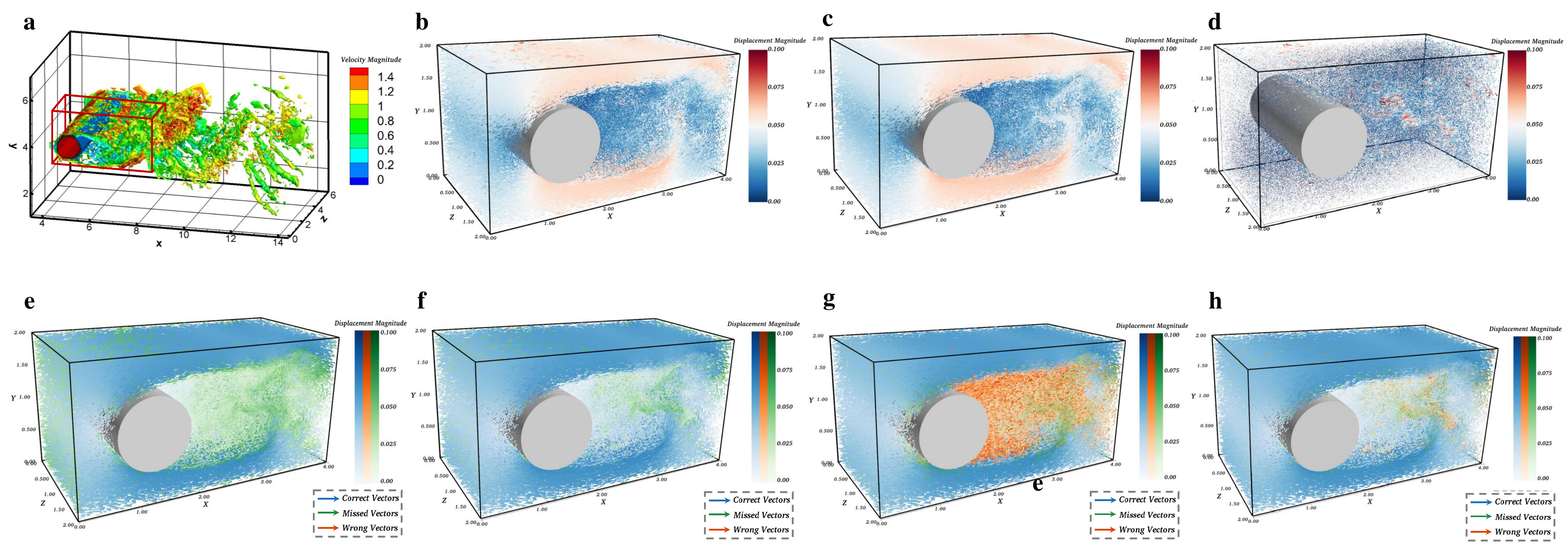}}
	\caption{\textbf{GotFlow3D applied for \textit{CylinderFlow} dataset, estimating the shedding flow motion over a cylinder. }  (a) Overview of the 3D vorticity iso-surfaces colored by velocity magnitude in the entire domain, with a specific subdomain outlined by a red box.  (b) Flow field in the subdomain estimated by GotFlow3D, where the color represents the displacement magnitude. (c) Ground-truth flow field, where the color represents the displacement magnitude. (d) Residual error between GotFlow3D and ground-truth. The EPE in this case is approximately 0.0058. (e-h) The results of different particle tracking algorithms, including (e) T-PT, (f) GotFlow3D+T-PT, (g) SerialTrack, (h) GotFlow3D+SerialTrack. Correct vectors (blue), missed vectors (green) and wrong vectors (orange) are shown in the plots. 
		\label{fig:cylinder}}
\end{figure*}

To visually validate the contribution of GotFlow3D, we plot the estimated cylinder flow as well as the ground-truth flow of the experiment 60. An overview of the flow field in a large domain is shown in Figure~\ref{fig:cylinder}(a), where a subdomain, that we mainly focus on, is outlined. 
As shown in Figure~\ref{fig:cylinder}(b), GotFlow3D itself has fulfilled an excellent recovery of the cylinder wake flow structure, which is nearly aligned with the ground truth (Figure~\ref{fig:cylinder}c). 
We also present the residual flow (i.e., the error between estimated flow and the ground truth) in Figure~\ref{fig:cylinder}(d), where the error mainly exists in a small region behind the cylinder. 
Furthermore, the tracking results of the two open-source PTV algorithms as well as their combinations with GotFlow3D are visually illustrated in Figure~\ref{fig:cylinder}(e-h).
We present the correct matches in blue vectors, missed matches in green vectors (i.e., the complementary set of correct matches in possible matches) and wrong matches in red vectors (i.e., the complementary set of correct matches in extracted matches), respectively. 
It can be seen that T-PT strives to maintain a high tracking reliability, and consequently misses plenty of potential matches, which however are successfully brought back by GotFlow3D.
On the contrary, SerialTrack focuses on the extraction of more potential matches, and therefore results in abundant wrong matches, which are also well corrected by GotFlow3D.
It is worth noting that although our synthetic training dataset cannot cover all flow scenarios (e.g., the cylinder flow), the impressive performance GotFlow3D makes it a promising method to be generalized to more potential and realistic applications.

\subsection{GotFlow3D for deformation of sphere indentation}

\begin{figure*}[t]
	\centering{\includegraphics[width=6.6in]{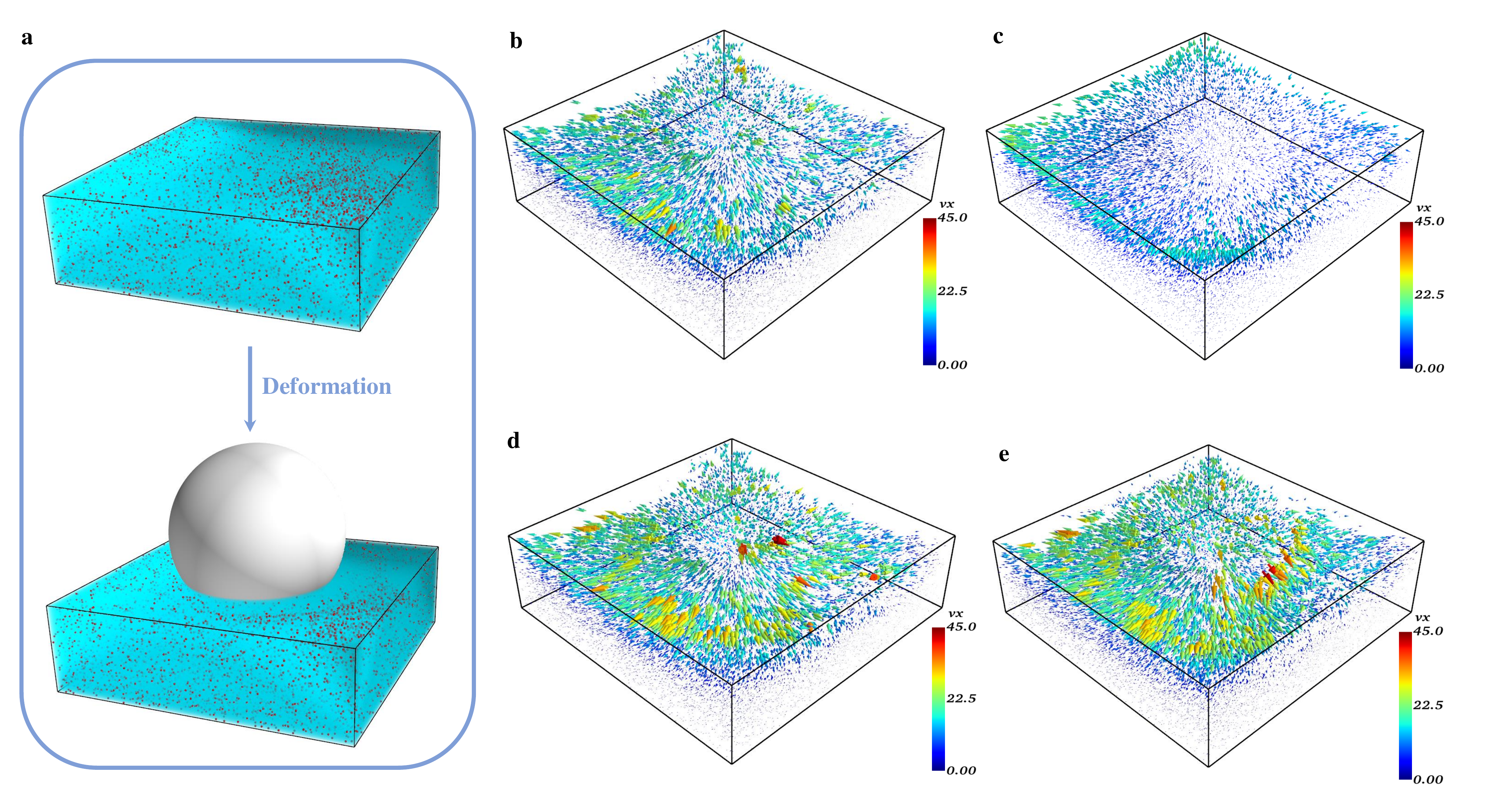}}
	\caption{\textbf{GotFlow3D applied for deformative flow of sphere indentation. } (a) Reference and deformed configuration of the sphere indentation, showing the deformation of the flow motion. (b-e) Tracking vector fields of (b) T-PT, (c) SerialTrack, (d) GotFlow3D+T-PT and (e) GotFlow3D+SerialTrack. With GotFlow3D combined, the tracking algorithms can extract more potential matches and demonstrate more flow structures. 
		\label{fig:appexp}}
\end{figure*}

To further investigate the generalization to real experiments, we apply GotFlow3D to recover the 3D spherical indentation deformation data \textit{DeformationFlow}.
An illustration of the reference and deformed experimental configurations is shown in Figure~\ref{fig:appexp}(a). 
About 35000 particles (plotted as red dots) are simultaneously traced in the investigated domain of 1024 voxels $\times$ 1024 voxels $\times$ 445 voxels.
The investigated  flow motion of the hydrogel deformation exhibits an approximate displacement ratio of $\rho = 0.49$ according to our estimation. 
The velocity vectors computed by T-PT, SerialTrack, the integrated ``GotFlow3D+T-PT'' and ``GotFlow3D+SerialTrack'' are presented in Figure~\ref{fig:appexp} for comparison. Parameters of SerialTrack are set as the recommended ones from the original paper \cite{yang2022serialtrack}.

The recovered  displacements of these methods are visualized in Figure~\ref{fig:appexp}(b-e).
It can be observed that ``GotFlow3D+T-PT'' (Figure~\ref{fig:appexp}d) and ``GotFlow3D+SerialTrack'' (Figure~\ref{fig:appexp}e) yield denser 3D flow fields than T-PT and SerialTrack (Figure~\ref{fig:appexp}a-b). 
The improvement of spatial resolution comes from the flow prediction of GotFlow3D, which contributes to covering more fine details of the revealed complex flow structures. With GotFlow3D combined, the tracking algorithms can extract more potential matches and demonstrate more flow structures. 
Moreover, the estimations of ``GotFlow3D+T-PT'' and ``GotFlow3D+SerialTrack'' are nearly consistent with the results of T-PT and SerialTrack in the hydrogel region of relatively small displacements, which is far away from the sphere and mainly distributed with blue vectors.
However, both original T-PT and SerialTrack fail to deal with the flow in the region near the contact surface between the sphere and the hydrogel. They provide sparse flow fields with small magnitudes, but relatively large particle displacements are expected. 
On the contrary, ``GotFlow3D+T-PT'' and ``GotFlow3D+SerialTrack'' show a great capability in extracting large deformation fields, especially in the contact surface region.
Through the comparison, we can conclude that GotFlow3D, which is trained using synthetic flow, can be effectively generalized to more complex flow scenarios in real-world experiments and show excellent performance in resolving large displacements in particle tracking.


\section{Discussion}


In this paper, we introduce a complete pipeline to employ the deep neural network called GotFlow3D to learn the 3D fluid motion from two consecutive unorganized particle sets. The contributions of this work are manifold. 
A static-dynamic fusion graph neural network (\textit{SDGNN}) is first proposed to derive the distinctive pointwise feature of particle sets by fusing the extracted geometric information from the static spatial graph and the dynamic feature graph. 
We present a novel approach to retrieve the adaptive point-voxel optimal transport (\textit{AdaPT}) plan for the iterative flow learning. The retrieval of the \textit{AdaPT} plan effectively provides the long-range and small-scale correspondence of particle pairs.
We generate the first 3D synthetic fluid flow dataset named \textit{FluidFlow3D} for the network training based on the physical simulations from CFD.
This dataset covers a great deal of complex non-rigid motion under different flow conditions beyond the common rigid motion in scene flow tasks. 
The generated dataset is available for  broader research in both the fluid mechanics community and the computer vision community.
Comprehensive evaluations on synthetic and real-world experimental data have shown that our approach rivals state-of-the-art scene flow networks and other advanced PTV methods.
In particular, our approach demonstrates competitive accuracy to the prior art, excellent robustness to various flow conditions and impressive generalization to real-world applications.


We devise a new framework to learn 3D complex fluid flow motion from consecutive particle sets in particle tracking experiments, which draws inspiration from some pioneering learning-based networks. However, there are obvious differences between GotFlow3D and the existing scene flow learners. 
The design of the proposed network only follows the basic similar concept of RAFT to estimate the flow in an iterative manner.
In fact, RAFT performs the flow estimation on regular images, while our GotFlow3D targets at irregular particle sets.
PV-RAFT adopts a similar iterative flow update to deal with point clouds, which however pays more attention to the basic point-voxel correlation fields.
In comparison, we  propose to adaptively retrieve a more efficient \textit{AdaPT} plan to boost the recurrent flow learning guided by optimal transport.
Both FLOT and GotFlow3D seek guidance from optimal transport theory, but implement that in different ways.
The former directly employs optimal transport to compute a coarse flow, whereas the latter performs the \textit{AdaPT} plan retrieval to extract more reliable correspondence information for the subsequent iterative flow refinement.
Apart from the above-demonstrated differences, the innovation of the proposed  GotFlow3D is also embodied in the design of the \textit{SDGNN}, which makes the first attempt to combine features from the dynamic feature graph and that of the static spatial graph in flow motion learning tasks.
Such combination of new modules presents a significant improvement on the accuracy of flow estimation according to the ablation study shown in Table~\ref{tab6:ablation}.


In summary, a novel deep learning based model - GotFlow3D - is proposed for learning complex flow motion from two particle sets, which  provides a new perspective for introducing deep learning approaches in the further analysis of particle tracking applications encountered in a wide range of biological and physical systems.
We present the first synthetic dataset for further deep learning developments in PTV community. GotFlow3D shows state-of-the-art accuracy on the  PTV database and surpasses the existing scene flow neural networks on most flow cases.
Robustness analysis further validates the stable performance of GotFlow3D under challenging flow cases and extreme experimental conditions, which are normally encountered in reality.
GotFlow3D also achieves a significant performance on the experiment of cylinder flow, which does not even exist in the training database, and brings a great improvement to the existing PTV approaches.
Moreover, GotFlow3D has been generalized to a real-world particle tracking application of sphere indentation, where  GotFlow3D shows the  superiority  of improving the spatial resolution and resolving large displacements.

Future extensions to deep learning approaches in PTV analysis include Bayesian framework of GotFlow3D, which could provide not only flow estimation but also the uncertainty quantification of the measurement. Moreover, GotFlow3D is currently designed as an universal tool, which simply consumes spatial coordinates of particles. More properties of particles (e.g., intensity) could be utilized to further improve the correspondence searching and make GotFlow3D customized tools in specific applications.
It is also a promising research topic to incorporate physics-informed  constraints (e.g., governing equations of the flow motion) in specific applications while training the neural networks.

\appendices

\renewcommand{\figurename}{Appendix Figure}
\renewcommand{\tablename}{Appendix Table}

\renewcommand\arraystretch{2.}
\begin{table*}[h]
	\centering
	\scriptsize
	\caption{\textbf{Configuration of various flow cases in the synthetic dataset.} We generate a training dataset with more than 16000 samples, each of which contains two particle sets and the corresponding ground truth flow. The synthetic dataset consists of six kinds of flow scenarios generated by using computational fluid dynamics (CFD), where some of them are extracted from the Johns Hopkins Turbulence Databases. Here, we list the name of flow motion patterns, the range of observation volume, the range of duration time for particle trajectories in the simulation and the range of displacement ratio $\rho$. By randomly selecting the parameters in the ranges for each data item, we generate a dataset mimicking diverse  conditions in PTV experiments. 
		\label{tab1:dataset}}
	\resizebox{0.75\textwidth}{!}
	{\begin{tabular*}{26pc}{@{\extracolsep{\fill}}llll@{}}
			\toprule[1pt]
			Flow case                      &\makecell[l]{Observation Volume}        &$\Delta t$  & $\rho$ \\
			\midrule
			\makecell[l]{Forced \\isotropic \\turbulence} &\makecell[l]{\{(0.25$\pi$, 0.25$\pi$, 0.25$\pi$), \\ \ (0.5$\pi$, 0.5$\pi$, 0.5$\pi$)\}}   & \{0.03, 0.04, 0.05\}    &[0.36, 2.50] \\
			\hline
			\makecell[l]{Forced \\MHD \\turbulence} &\makecell[l]{\{(0.25$\pi$, 0.25$\pi$, 0.25$\pi$), \\ \ (0.5$\pi$, 0.5$\pi$, 0.5$\pi$)\}}   & \{0.06, 0.07, 0.08\}     &[0.58, 2.33] \\
			\hline
			\makecell[l]{Transitional \\boundary \\flow} &\makecell[l]{\{(0.5$\pi$, 1, 0.5$\pi$) , \\ \ ($\pi$, 2, $\pi$)\}}   & \{0.12, 0.16, 0.20\}     &[0.48, 2.24] \\
			\hline
			\makecell[l]{Turbulent \\channel \\flow} &(0.25$\pi$, 0.5, 0.25$\pi$)   & \makecell[l]{\{0.03, 0.04, 0.05, \\ 0.06, 0.07, 0.08\}}     &[0.51, 1.63] \\
			\hline
			\makecell[l]{Beltrami \\flow} &(2, 2, 2)   & \makecell[l]{\{ [0.9, 1.0], \\ \ [1.1, 1.2], \\ \ [1.3, 1.4] \}}    &[0.35, 6.58] \\
			\hline
			\makecell[l]{Uniform \\flow} &(0.25$\pi$, 0.25$\pi$, 0.25$\pi$)   & \{0.03, 0.04, 0.05\}     &[0.51, 3.11] \\                                                               	      
			\bottomrule[1pt]
	\end{tabular*}}{}
\end{table*}

\renewcommand\arraystretch{1.2}
\begin{table*}[h]
	\centering
	\caption{\textbf{Performance comparison between the state-of-the-art scene flow learning methods and GotFlow3D on different flow cases in the synthetic dataset.} The best results are marked in bold.  We present the performance of these methods with metrics of EPE, ACC Strict, ACC Relax and Outliers. GotFlow3D achieves the best performance on most of the flow cases except the Uniform flow, where PV-RAFT is slightly superior to GotFlow3D on ACC Strict (Relax) as well as Outliers. In some of the complex flow cases (e.g., Forced isotropic turbulence, Forced MHD turbulence and Beltrami flow) which contain small-scale flow structures, the existing flow estimators including PV-RAFT provide worse results, whereas GotFlow3D consistently performs the best.   \label{tab3:flowcase}}
	\resizebox{\textwidth}{!}
	{\begin{tabular*}{66pc}{@{\extracolsep{\fill}}c|cccc|cccc|cccc@{}}
			\toprule[1pt]
			\multirow{2}{*}{\makecell*[c]{\\ Method}}   &\multicolumn{4}{c|}{\makecell[c]{Forced isotropic turbulence}}      &\multicolumn{4}{c|}{\makecell[c]{Forced MHD turbulence}} & \multicolumn{4}{c}{\makecell[c]{Transitional boundary flow}}  \\
			\cmidrule(lr){2-5} \cmidrule(lr){6-9} \cmidrule(lr){10-13} 
			&EPE &\makecell[c]{Acc \\ Strict}  & \makecell[c]{Acc \\ Relax} & Outliers &EPE &\makecell[c]{Acc \\ Strict}  & \makecell[c]{Acc \\ Relax} & Outliers &EPE &\makecell[c]{Acc \\ Strict}  & \makecell[c]{Acc \\ Relax} & Outliers \\
			\midrule
			FlowNet3D &0.12465   &0.20\%   &1.52\% &98.48\%  &0.09400 &0.06\% &0.45\%  &99.55\% &0.02003 &67.23\%  &91.01\% &8.99\% \\ 
			FLOT  &0.13090   &0.14\%   &1.15\% &98.85\% &0.09842 &0.04\% &0.28\%  &99.72\% &0.01715 &70.37\%  &94.63\% &5.37\% \\  
			PointPWC-Net &0.02719   &18.47\% &57.06\%  &42.94\% &0.01714 &8.46\% &33.53\%  &66.47\% &0.01163 &89.13\%  &98.29\% &1.71\%\\  
			PV-RAFT &0.04190   &51.37\%   &62.82\% &37.18\% &0.02594 &41.44\% &61.37\%  &38.63\% &0.00324 &99.65\%  &99.94\% &0.063\% \\  
			GotFlow3D     &\textbf{0.01153}   &\textbf{86.62\%}   &\textbf{90.74\%} &\textbf{9.26\%} &\textbf{0.00596} &\textbf{83.005\%} &\textbf{91.81\%}  &\textbf{8.19\%} &\textbf{0.00222} &\textbf{99.87\%}  &\textbf{99.97\%} &\textbf{0.028\%} \\    
			\bottomrule[1pt]
			\toprule[1pt]
			\multirow{2}{*}{\makecell*[c]{\\ Method}}   & \multicolumn{4}{c|}{\makecell[c]{Turbulent channel flow}} &  \multicolumn{4}{c|}{\makecell[c]{Beltrami flow}} & \multicolumn{4}{c}{\makecell[c]{Uniform flow}}  \\
			\cmidrule(lr){2-5} \cmidrule(lr){6-9} \cmidrule(lr){10-13} 
			&EPE &\makecell[c]{Acc \\ Strict}  & \makecell[c]{Acc \\ Relax} & Outliers &EPE &\makecell[c]{Acc \\ Strict}  & \makecell[c]{Acc \\ Relax} & Outliers &EPE &\makecell[c]{Acc \\ Strict}  & \makecell[c]{Acc \\ Relax} & Outliers \\
			\midrule
			FlowNet3D &0.03670   &24.89\%   &70.78\% &29.22\%  &0.04435 &2.30\% &13.73\%  &86.27\% &0.03748 &25.41\%  &79.23\% &20.77\% \\
			FLOT &0.03541   &26.89\%   &72.72\% &27.28\% &0.02417 &13.11\% &42.22\%  &57.78\% &0.02059 &68.25\%  &96.45\% &3.55\% \\  
			PointPWC-Net &0.01266   &87.51\% &98.41\%  &1.59\% &0.01616 &24.36\% &62.19\%  &37.81\% &0.02140 &64.31\%  &97.32\% &2.68\% \\   
			PV-RAFT  &0.00653 &93.71\% &97.83\%  &2.17\% &0.00899 &69.69\% &91.00\%  &9.00\% &0.00316 &\textbf{99.87\%}  &\textbf{99.99\%} &\textbf{0.015\%} \\  
			GotFlow3D    &\textbf{0.00241} &\textbf{99.13\%} &\textbf{99.74\%}  &\textbf{0.26\%} &\textbf{0.00291}   &\textbf{95.03\%}   &\textbf{98.44\%} &\textbf{1.56\%} &\textbf{0.00244} &99.85\%  &99.98\% &0.019\% \\        	      
			\bottomrule[1pt]
	\end{tabular*}}{}
\end{table*}

\renewcommand\arraystretch{1.2}
\begin{table*}[!htb]
	\centering
	\caption{\textbf{Ablation study on different modules in GotFlow3D}. We conduct a set of ablation experiments to demonstrate the effectiveness of each module in GotFlow3D, including the static and/or dynamic graph neural network for feature extraction, optimal transport and/or \textit{AdaPT} module for correspondence retrieval.  A green tick means the corresponding module is used, while a red cross means not used.   It can be seen from Networks 5-7 that the proposed dynamic GNN greatly improves the performance and the fusion of two kinds of graph (i.e., SDGNN ) further boosts the accuracy to a higher level.  One can also observe from Networks 1-2 that the replacement of correlation fields with the optimal transport reduces the error to some extent by modeling the correspondence searching with the transport plan optimization. As shown by the results of Networks 2-5, both proposed adaptive transport plan $\mathcal{T}^*_p$ and $\mathcal{T}^*_v$ are better than the point-voxel based lookup, and the integration of the $\mathcal{T}^*_p$ and $\mathcal{T}^*_v$ (Netwrok 5) performs the best. 
		\label{tab6:ablation}}
	\resizebox{\textwidth}{!}
	{\begin{tabular*}{58pc}{@{\extracolsep{\fill}}c|ccc|ccc|ccc|cc@{}}
			\toprule[1pt]
			\multirow{4}{*}{\makecell*[c]{\\ \\ Method}}    &\multicolumn{9}{c|}{Module options} &\multicolumn{2}{c}{ Metrics } \\
			\cmidrule(lr){2-10} \cmidrule(lr){11-12}
			&\multicolumn{3}{c|}{Feature Extraction} &\multicolumn{6}{c|}{Correspondence Retrieval} &\multirow{3}{*}{\makecell*[c]{\\ EPE}} & \multirow{3}{*}{\makecell[c]{\\ Outliers}}\\
			\cmidrule(lr){2-4} \cmidrule(lr){5-10}
			&\multirow{2}{*}{\makecell*[c]{Static \\Graph}} &\multirow{2}{*}{\makecell*[c]{Dynamic \\Graph}} &\multirow{2}{*}{\textit{SDGNN}} &\multirow{2}{*}{\makecell*[c]{Correlation \\Fields}} &\multirow{2}{*}{\makecell*[c]{Optimal \\Transport}} &\multirow{2}{*}{\makecell*[c]{Point-Voxel \\Lookup}} &\multicolumn{3}{c|}{\textit{AdaPT}} && \\
			\cmidrule(lr){8-10}
			& & & & & & &$\mathcal{T}^*_p$+$\mathcal{T}_v$ &$\mathcal{T}_p$+$\mathcal{T}^*_v$ &$\mathcal{T}^*_p$+$\mathcal{T}^*_v$ & &  \\
			\midrule
			Network1 &\textcolor{c1}{\ding{52}} &\color{red}\XSolidBrush &\color{red}\XSolidBrush &\textcolor{c1}{\ding{52}} &\color{red}\XSolidBrush &\textcolor{c1}{\ding{52}} &\color{red}\XSolidBrush &\color{red}\XSolidBrush &\color{red}\XSolidBrush  &0.01651  &16.39\% \\
			Network2 &\textcolor{c1}{\ding{52}} &\color{red}\XSolidBrush &\color{red}\XSolidBrush  &\color{red}\XSolidBrush &\textcolor{c1}{\ding{52}}  &\textcolor{c1}{\ding{52}}   &\color{red}\XSolidBrush &\color{red}\XSolidBrush &\color{red}\XSolidBrush  &0.01208  &10.24\% \\        
			Network3 &\textcolor{c1}{\ding{52}} &\color{red}\XSolidBrush &\color{red}\XSolidBrush &\color{red}\XSolidBrush &\textcolor{c1}{\ding{52}} &\color{red}\XSolidBrush   &\textcolor{c1}{\ding{52}}   &\color{red}\XSolidBrush &\color{red}\XSolidBrush &0.01177  &9.85\% \\
			Network4 &\textcolor{c1}{\ding{52}} &\color{red}\XSolidBrush &\color{red}\XSolidBrush &\color{red}\XSolidBrush &\textcolor{c1}{\ding{52}} &\color{red}\XSolidBrush   &\color{red}\XSolidBrush   &\textcolor{c1}{\ding{52}} &\color{red}\XSolidBrush &0.01036  &9.43\% \\
			Network5 &\textcolor{c1}{\ding{52}} &\color{red}\XSolidBrush &\color{red}\XSolidBrush &\color{red}\XSolidBrush &\textcolor{c1}{\ding{52}} &\color{red}\XSolidBrush   &\color{red}\XSolidBrush   &\color{red}\XSolidBrush &\textcolor{c1}{\ding{52}} &0.00916  &7.96\% \\  
			Network6 &\color{red}\XSolidBrush &\textcolor{c1}{\ding{52}} &\color{red}\XSolidBrush &\color{red}\XSolidBrush &\textcolor{c1}{\ding{52}} &\color{red}\XSolidBrush   &\color{red}\XSolidBrush   &\color{red}\XSolidBrush &\textcolor{c1}{\ding{52}} &0.00603  &4.46\% \\  
			Network7 &\color{red}\XSolidBrush &\color{red}\XSolidBrush &\textcolor{c1}{\ding{52}} &\color{red}\XSolidBrush &\textcolor{c1}{\ding{52}} &\color{red}\XSolidBrush   &\color{red}\XSolidBrush   &\color{red}\XSolidBrush &\textcolor{c1}{\ding{52}} &\textbf{0.00487}  &\textbf{3.63\%} \\           	      
			\bottomrule[1pt]
	\end{tabular*}}{}
\end{table*}

\ifCLASSOPTIONcompsoc
  \section*{Acknowledgments}
\else
  \section*{Acknowledgment}
\fi

This work was supported in part by the National Key R\&D Program of China (No. 2019YFB1705800), in part by the National Natural Science Foundation of China under grant No.~61973270, in part by the Foundation for Innovative Research Groups of the National Natural Science Foundation of China under grant No.~61621002.

\ifCLASSOPTIONcaptionsoff
  \newpage
\fi



%
%
%

\normalem
\bibliographystyle{IEEEtran}
\bibliography{IEEEabrv,refe_ljm.bib}

\begin{thebibliography}{10}
\providecommand{\url}[1]{#1}
\csname url@samestyle\endcsname
\providecommand{\newblock}{\relax}
\providecommand{\bibinfo}[2]{#2}
\providecommand{\BIBentrySTDinterwordspacing}{\spaceskip=0pt\relax}
\providecommand{\BIBentryALTinterwordstretchfactor}{4}
\providecommand{\BIBentryALTinterwordspacing}{\spaceskip=\fontdimen2\font plus
\BIBentryALTinterwordstretchfactor\fontdimen3\font minus
  \fontdimen4\font\relax}
\providecommand{\BIBforeignlanguage}[2]{{%
\expandafter\ifx\csname l@#1\endcsname\relax
\typeout{** WARNING: IEEEtran.bst: No hyphenation pattern has been}%
\typeout{** loaded for the language `#1'. Using the pattern for}%
\typeout{** the default language instead.}%
\else
\language=\csname l@#1\endcsname
\fi
#2}}
\providecommand{\BIBdecl}{\relax}
\BIBdecl

\bibitem{kemp2019leonardo}
M.~Kemp, ``{Leonardo da Vinci's} laboratory: studies in flow,'' \emph{Nature},
  vol. 571, no. 7765, pp. 322--324, 2019.

\bibitem{dabiri2020particle}
D.~Dabiri and C.~Pecora, \emph{{Particle Tracking Velocimetry}}.\hskip 1em plus
  0.5em minus 0.4em\relax IOP Publishing Bristol, 2020.

\bibitem{kopitca2021programmable}
A.~Kopitca, K.~Latifi, and Q.~Zhou, ``Programmable assembly of particles on a
  {Chladni} plate,'' \emph{Science advances}, vol.~7, no.~39, p. eabi7716,
  2021.

\bibitem{ferdowsi2017river}
B.~Ferdowsi, C.~P. Ortiz, M.~Houssais, and D.~J. Jerolmack, ``River-bed
  armouring as a granular segregation phenomenon,'' \emph{Nature
  communications}, vol.~8, no.~1, pp. 1--10, 2017.

\bibitem{hu2003hydrodynamics}
D.~L. Hu, B.~Chan, and J.~W. Bush, ``The hydrodynamics of water strider
  locomotion,'' \emph{Nature}, vol. 424, no. 6949, pp. 663--666, 2003.

\bibitem{he2014apical}
B.~He, K.~Doubrovinski, O.~Polyakov, and E.~Wieschaus, ``Apical constriction
  drives tissue-scale hydrodynamic flow to mediate cell elongation,''
  \emph{Nature}, vol. 508, no. 7496, pp. 392--396, 2014.

\bibitem{mestre2020cerebrospinal}
H.~Mestre, T.~Du, A.~M. Sweeney, G.~Liu, A.~J. Samson, W.~Peng, K.~N.
  Mortensen, F.~F. St{\ae}ger, P.~A. Bork, L.~Bashford \emph{et~al.},
  ``Cerebrospinal fluid influx drives acute ischemic tissue swelling,''
  \emph{Science}, vol. 367, no. 6483, p. eaax7171, 2020.

\bibitem{zhang2022cerebral}
Z.~Zhang, M.~Hwang, T.~J. Kilbaugh, A.~Sridharan, and J.~Katz, ``Cerebral
  microcirculation mapped by echo particle tracking velocimetry quantifies the
  intracranial pressure and detects ischemia,'' \emph{Nature communications},
  vol.~13, no.~1, pp. 1--15, 2022.

\bibitem{guo2014probing}
M.~Guo, A.~J. Ehrlicher, M.~H. Jensen, M.~Renz, J.~R. Moore, R.~D. Goldman,
  J.~Lippincott-Schwartz, F.~C. Mackintosh, and D.~A. Weitz, ``Probing the
  stochastic, motor-driven properties of the cytoplasm using force spectrum
  microscopy,'' \emph{Cell}, vol. 158, no.~4, pp. 822--832, 2014.

\bibitem{peng2021imaging}
Y.~Peng, Z.~Liu, and X.~Cheng, ``Imaging the emergence of bacterial turbulence:
  {Phase} diagram and transition kinetics,'' \emph{Science advances}, vol.~7,
  no.~17, p. eabd1240, 2021.

\bibitem{schuerle2019synthetic}
S.~Schuerle, A.~P. Soleimany, T.~Yeh, G.~Anand, M.~H{\"a}berli, H.~Fleming,
  N.~Mirkhani, F.~Qiu, S.~Hauert, X.~Wang \emph{et~al.}, ``Synthetic and living
  micropropellers for convection-enhanced nanoparticle transport,''
  \emph{Science advances}, vol.~5, no.~4, p. eaav4803, 2019.

\bibitem{punzmann2014generation}
H.~Punzmann, N.~Francois, H.~Xia, G.~Falkovich, and M.~Shats, ``Generation and
  reversal of surface flows by propagating waves,'' \emph{Nature physics},
  vol.~10, no.~9, pp. 658--663, 2014.

\bibitem{huang2013imaging}
P.~Y. Huang, S.~Kurasch, J.~S. Alden, A.~Shekhawat, A.~A. Alemi, P.~L. McEuen,
  J.~P. Sethna, U.~Kaiser, and D.~A. Muller, ``Imaging atomic rearrangements in
  two-dimensional silica glass: watching silica’s dance,'' \emph{Science},
  vol. 342, no. 6155, pp. 224--227, 2013.

\bibitem{pereira2006two}
F.~Pereira, H.~St{\"u}er, E.~C. Graff, and M.~Gharib, ``Two-frame {3D} particle
  tracking,'' \emph{Measurement science and technology}, vol.~17, no.~7, p.
  1680, 2006.

\bibitem{leggett2020mechanophenotyping}
S.~E. Leggett, M.~Patel, T.~M. Valentin, L.~Gamboa, A.~S. Khoo, E.~K. Williams,
  C.~Franck, and I.~Y. Wong, ``Mechanophenotyping of {3D} multicellular
  clusters using displacement arrays of rendered tractions,'' \emph{Proceedings
  of the National Academy of Sciences}, vol. 117, no.~11, pp. 5655--5663, 2020.

\bibitem{raffel2018particle}
M.~Raffel, C.~Willert, F.~Scarano, C.~K{\"a}hler, S.~Wereley, and
  J.~Kompenhans, \emph{{Particle Image Velocimetry: A Practical Guide}}.\hskip
  1em plus 0.5em minus 0.4em\relax Cham: Springer International Publishing,
  2018.

\bibitem{cierpka2013higher}
C.~Cierpka, B.~L{\"u}tke, and C.~J. K{\"a}hler, ``Higher order multi-frame
  particle tracking velocimetry,'' \emph{Experiments in Fluids}, vol.~54,
  no.~5, pp. 1--12, 2013.

\bibitem{brunton2020machine}
S.~L. Brunton, B.~R. Noack, and P.~Koumoutsakos, ``Machine learning for fluid
  mechanics,'' \emph{Annual Review of Fluid Mechanics}, vol.~52, pp. 477--508,
  2020.

\bibitem{cai2019dense}
S.~Cai, S.~Zhou, C.~Xu, and Q.~Gao, ``Dense motion estimation of particle
  images via a convolutional neural network,'' \emph{Experiments in Fluids},
  vol.~60, no.~4, pp. 1--16, 2019.

\bibitem{cai2019particle}
S.~Cai, J.~Liang, Q.~Gao, C.~Xu, and R.~Wei, ``Particle image velocimetry based
  on a deep learning motion estimator,'' \emph{IEEE Transactions on
  Instrumentation and Measurement}, vol.~69, no.~6, pp. 3538--3554, 2019.

\bibitem{liang2020filtering}
J.~Liang, S.~Cai, C.~Xu, and J.~Chu, ``Filtering enhanced tomographic {PIV}
  reconstruction based on deep neural networks,'' \emph{IET Cyber-Systems and
  Robotics}, vol.~2, no.~1, pp. 43--52, 2020.

\bibitem{lagemann2021deep}
C.~Lagemann, K.~Lagemann, S.~Mukherjee, and W.~Schr{\"o}der, ``Deep recurrent
  optical flow learning for particle image velocimetry data,'' \emph{Nature
  Machine Intelligence}, vol.~3, no.~7, pp. 641--651, 2021.

\bibitem{mallery2020dense}
K.~Mallery, S.~Shao, and J.~Hong, ``Dense particle tracking using a learned
  predictive model,'' \emph{Experiments in Fluids}, vol.~61, no.~10, pp. 1--14,
  2020.

\bibitem{qi2017pointnet}
C.~R. Qi, H.~Su, K.~Mo, and L.~J. Guibas, ``Pointnet: {Deep} learning on point
  sets for 3d classification and segmentation,'' in \emph{Proceedings of the
  IEEE conference on computer vision and pattern recognition}, 2017, pp.
  652--660.

\bibitem{liu2019flownet3d}
X.~Liu, C.~R. Qi, and L.~J. Guibas, ``{Flownet3D}: Learning scene flow in {3D}
  point clouds,'' in \emph{Proceedings of the IEEE/CVF Conference on Computer
  Vision and Pattern Recognition}, 2019, pp. 529--537.

\bibitem{puy2020flot}
G.~Puy, A.~Boulch, and R.~Marlet, ``Flot: {Scene} flow on point clouds guided
  by optimal transport,'' in \emph{European conference on computer
  vision}.\hskip 1em plus 0.5em minus 0.4em\relax Springer, 2020, pp. 527--544.

\bibitem{wu2020pointpwc}
W.~Wu, Z.~Y. Wang, Z.~Li, W.~Liu, and L.~Fuxin, ``{PointPWC-Net}: {Cost} volume
  on point clouds for (self-) supervised scene flow estimation,'' in
  \emph{European conference on computer vision}.\hskip 1em plus 0.5em minus
  0.4em\relax Springer, 2020, pp. 88--107.

\bibitem{wei2021pv}
Y.~Wei, Z.~Wang, Y.~Rao, J.~Lu, and J.~Zhou, ``{PV-RAFT}: point-voxel
  correlation fields for scene flow estimation of point clouds,'' in
  \emph{Proceedings of the IEEE/CVF Conference on Computer Vision and Pattern
  Recognition}, 2021, pp. 6954--6963.

\bibitem{liang2021deepptv}
J.~Liang, S.~Cai, C.~Xu, T.~Chen, and J.~Chu, ``{DeepPTV: Particle Tracking
  Velocimetry} for complex flow mmtion via deep neural networks,'' \emph{IEEE
  Transactions on Instrumentation and Measurement}, 2021.

\bibitem{yang2022serialtrack}
J.~Yang, Y.~Yin, A.~K. Landauer, S.~Buyuktozturk, J.~Zhang, L.~Summey,
  A.~McGhee, M.~K. Fu, J.~O. Dabiri, and C.~Franck, ``{SerialTrack}: scale and
  rotation invariant augmented lagrangian particle tracking,'' \emph{arXiv
  preprint arXiv:2203.12573}, 2022.

\bibitem{qi2017pointnet++}
C.~R. Qi, L.~Yi, H.~Su, and L.~J. Guibas, ``Pointnet++: {Deep} hierarchical
  feature learning on point sets in a metric space,'' \emph{Advances in neural
  information processing systems}, vol.~30, 2017.

\bibitem{wang2019dynamic}
Y.~Wang, Y.~Sun, Z.~Liu, S.~E. Sarma, M.~M. Bronstein, and J.~M. Solomon,
  ``Dynamic graph cnn for learning on point clouds,'' \emph{Acm Transactions On
  Graphics (tog)}, vol.~38, no.~5, pp. 1--12, 2019.

\bibitem{villani2009optimal}
C.~Villani, \emph{Optimal transport: old and new}.\hskip 1em plus 0.5em minus
  0.4em\relax Springer, 2009, vol. 338.

\bibitem{peyre2019computational}
G.~Peyr{\'e}, M.~Cuturi \emph{et~al.}, ``Computational optimal transport: With
  applications to data science,'' \emph{Foundations and Trends{\textregistered}
  in Machine Learning}, vol.~11, no. 5-6, pp. 355--607, 2019.

\bibitem{cuturi2013sinkhorn}
M.~Cuturi, ``Sinkhorn distances: lightspeed computation of optimal transport,''
  \emph{Advances in neural information processing systems}, vol.~26, 2013.

\bibitem{teed2020raft}
Z.~Teed and J.~Deng, ``Raft: {Recurrent} all-pairs field transforms for optical
  flow,'' in \emph{European conference on computer vision}.\hskip 1em plus
  0.5em minus 0.4em\relax Springer, 2020, pp. 402--419.

\bibitem{kingma2014adam}
D.~P. Kingma and J.~Ba, ``{Adam}: {A} method for stochastic optimization,''
  \emph{arXiv preprint arXiv:1412.6980}, 2014.

\bibitem{li2008public}
Y.~Li, E.~Perlman, M.~Wan, Y.~Yang, C.~Meneveau, R.~Burns, S.~Chen, A.~Szalay,
  and G.~Eyink, ``A public turbulence database cluster and applications to
  study lagrangian evolution of velocity increments in turbulence,''
  \emph{Journal of Turbulence}, no.~9, p. N31, 2008.

\bibitem{ethier1994exact}
C.~R. Ethier and D.~Steinman, ``Exact fully {3D Navier--Stokes} solutions for
  benchmarking,'' \emph{International Journal for Numerical Methods in Fluids},
  vol.~19, no.~5, pp. 369--375, 1994.

\bibitem{maas1993particle}
H.~Maas, A.~Gruen, and D.~Papantoniou, ``Particle tracking velocimetry in
  three-dimensional flows,'' \emph{Experiments in fluids}, vol.~15, no.~2, pp.
  133--146, 1993.

\bibitem{khojasteh2022lagrangian}
A.~R. Khojasteh, S.~Laizet, D.~Heitz, and Y.~Yang, ``{Lagrangian and Eulerian}
  dataset of the wake downstream of a smooth cylinder at a {Reynolds} number
  equal to 3900,'' \emph{Data in brief}, vol.~40, p. 107725, 2022.

\bibitem{patel2018rapid}
M.~Patel, S.~E. Leggett, A.~K. Landauer, I.~Y. Wong, and C.~Franck, ``Rapid,
  topology-based particle tracking for high-resolution measurements of large
  complex {3D} motion fields,'' \emph{Scientific reports}, vol.~8, no.~1, pp.
  1--14, 2018.

\bibitem{scimone2018modular}
M.~T. Scimone, H.~C. Cramer~III, E.~Bar-Kochba, R.~Amezcua, J.~B. Estrada, and
  C.~Franck, ``Modular approach for resolving and mapping complex neural and
  other cellular structures and their associated deformation fields in three
  dimensions,'' \emph{Nature protocols}, vol.~13, no.~12, pp. 3042--3064, 2018.

\end{thebibliography}

%

%






\end{document}